\documentclass[fleqn,10pt]{wlscirep}
\usepackage[utf8]{inputenc}
\usepackage[T1]{fontenc}
\usepackage{algorithm}
\usepackage{algorithmic}
\title{Data re-uploading in Quantum Machine Learning for time series: application to traffic forecasting}

\author[1,2,*]{Nikolaos Schetakis}
\author[3]{Paolo Bonfini}
\author[4]{Negin Alisoltani}
\author[5]{Konstantinos Blazakis}
\author[6]{Symeon I. Tsintzos}
\author[6]{Alexis Askitopoulos}
\author[7]{Davit Aghamalyan}
\author[8]{Panagiotis Fafoutellis}
\author[8]{Eleni I. Vlahogianni}
\affil[1]{Computational Mechanics and Optimization Laboratory, School of Production Engineering and Management, Technical University of Crete, 73100 Chania, Greece}
\affil[2]{Quantum Innovation Pc, 73100 Chania, Greece}
\affil[3]{Alma-Sistemi Srl, 00012 Guidonia, Italy}
\affil[4]{Univ Gustave Eiffel, COSYS, GRETTIA, Paris, F-77454, France}
\affil[5]{School of Electrical and Computer Engineering, Technical University of Crete, 73100, Greece}
\affil[6]{QUBITECH Pc, Athens, 15231, Greece}
\affil[7]{Institute of High Performance Computing, A*STAR (Agency for Science, Technology and Research), 1 Fusionopolis Way, \#16-16 Connexis, Singapore 138632}
\affil[8]{Department of Transportation Planning and Engineering, School of Civil Engineering, National Technical University of Athens, 5 Iroon Polytechniou Str, Zografou Campus, 157 73 Athens, Greece}

\affil[*]{nischetakis@tuc.gr}

\begin{abstract}
Accurate traffic forecasting plays a crucial role in modern Intelligent Transportation Systems (ITS), as it enables real-time traffic flow management, reduces congestion, and improves the overall efficiency of urban transportation networks. With the rise of Quantum Machine Learning (QML), it has emerged a new paradigm possessing the potential to enhance predictive capabilities beyond what classical machine learning models can achieve.
In the present work we pursue a heuristic approach to explore the potential of QML, and focus on a specific transport issue.
In particular, as a case study we investigate a traffic forecast task for a major urban area in Athens (Greece), for which we possess high-resolution data.
In this endeavor we explore the application of Quantum Neural Networks (QNN), and, notably, we present the first application of quantum data re-uploading in the context of transport forecasting.  This technique allows quantum models to better capture complex patterns, such as traffic dynamics, by repeatedly encoding classical data into a quantum state.
Aside from providing a prediction model, we spend considerable effort in comparing the performance of our hybrid quantum-classical neural networks with classical deep learning approaches.
Our results show that hybrid models achieve competitive accuracy with state-of-the-art classical methods, especially when the number of qubits and re-uploading blocks is increased. While the classical models demonstrate lower computational demands, we provide evidence that increasing the complexity of the quantum model improves predictive accuracy. These findings indicate that QML techniques, and specifically the data re-uploading approach, hold promise for advancing traffic forecasting models and could be instrumental in addressing challenges inherent in ITS environments.

\end{abstract}
\begin{document}

\flushbottom
\maketitle
%
%
\thispagestyle{empty}


\section{Introduction}

Traffic forecasting is a crucial component of modern urban transportation systems, directly influencing traffic management, congestion control, and the efficiency of ITS \cite{Yin:2021, Lana:2018}. Accurate traffic predictions enable timely interventions to prevent traffic congestion, optimize traffic signal timing, and improve road utilization \cite{Cheng:2020, Gao:2022, Vlahogianni:2014, Yao:2019}. The increasing availability of high-resolution traffic data generated from sensors, connected vehicles, and other sources has led to the widespread adoption of Deep Learning (DL) models such as Long Short-Term Memory (LSTM) networks and Convolutional Neural Networks (CNNs) to address this complex forecasting task \cite{Mantouka:2021, Vlahogianni:2017}. These models have shown significant improvements in accuracy over traditional statistical and machine learning methods but come with limitations, particularly in computational efficiency and the ability to explain intricate relationships within the data \cite{Jiang:2021}.

A key challenge associated with these DL models lies in their computational demands, which include long training times, high complexity, and difficulties in capturing both the temporal and spatial dependencies inherent in traffic data. As transportation networks grow increasingly complex and the need for real-time forecasting becomes more critical, the search for more powerful and efficient alternatives to classical DL models intensifies. This raises an important question: can emerging quantum technologies help overcome these limitations? 

In the following section, we explore the potential advantages offered by these quantum approaches within the current era of quantum devices, known as the ``Noisy Intermediate-Scale Quantum'' (NISQ)\cite{bharti2022noisy,preskill2018quantum} era. This era is characterized by quantum processors with a moderate number of qubits (50–500), which, while promising, are still prone to noise and lack scalability, robust error correction, and the capability for full-scale quantum advantage.


\noindent
In the last few years, tremendous advancements have reshaped the NISQ landscape. NISQ devices have provided a glimpse into the potential of quantum systems to outperform classical computers in specific tasks. One of the landmark achievements in this domain was Google's demonstration of ``quantum supremacy'' using their Sycamore chip\cite{arute2019quantum}.
The term ``quantum supremacy''\cite{harrow2017quantum} refers to the ability of a quantum computer to perform a computation that is exponentially faster than any classical computer. However, Google's demonstration, while a significant milestone, was critiqued for not providing a practical advantage, as the specific task performed did not have direct real-world applications.
Another significant experiment in the realm of quantum advantage was conducted using the Jiuzhang photonic quantum computer\cite{zhong2020quantum}.
The quantum advantage was demonstrated in the complexity of sampling from a Torontonian matrix, which scales exponentially with the number of output photon clicks. Despite showing a quantum advantage, the Jiuzhang experiment did not achieve quantum supremacy because the photonic quantum computer used was not programmable.

It is truly breathtaking how rapid our experimental progress is in paving the way toward fault-tolerant error correction. On this journey first, a pioneering contribution arrived in 2022 from M. Lukin's group\cite{bluvstein2022quantum}, where they utilized a 2-dimensional controllable atomic array of 256 Rydberg atoms in order to achieve faithful error mitigation. This experiment equipped the community with the confidence that we might not be that far behind to demonstrate an example in the current quantum computers that error correction can work. Such a demonstration would imply the error threshold theorem, which in turn implies that with an increasing number of qubits, errors decrease in the system.
With this kind of target in mind, researchers at Google pioneered its Willow 100 qubit quantum chip\cite{bravyi2024high} and recently achieved error correction below the error threshold.
Given this experimental progress, which has happened in just the last few years, one might argue that we are slowly entering into the \textit{post-}NISQ era, where, by building capabilities in error-correction, error-mitigation, and dynamical noise suppression, we will eventually lead to the arrival of the fully \textit{fault-tolerant} era.
Thus, the current quantum computing era might be more appropriately called the "soft post-NISQ" era.

It has been well conceived by several experts in the field that one of the most promising areas of research to obtain practical advantage is Quantum Machine Learning (QML)\cite{biamonte2017quantum,cerezo2022challenges,bowles2024better}, which has emerged through the cross-fertilization of ideas and methodologies between Quantum Computing and Classical Machine Learning. QML has indeed become a fascinating research area, offering potential quantum speedups for various computational tasks.
For example, the HHL algorithm\cite{harrow2009quantum}, named after its developers Harrow, Hassidim, and Lloyd, is a well-known quantum algorithm for solving linear systems of equations and is often cited as an example of quantum speedup. It provides exponential speedup in specific scenarios compared to classical algorithms, particularly when dealing with sparse matrices or when only some properties of the solution are required \cite{huang2019near}. Due to the growing interest in this topic, the latest developments are now covered in recent surveys  \cite{morales2024quantum}.
Similarly, quantum algorithms for tasks like finding eigenvectors and eigenvalues\cite{rebentrost2018quantum}, as well as performing principal component analysis (PCA)\cite{lloyd2014quantum}, have shown promise in achieving speedups over classical methods.

These algorithms use quantum parallelism and other quantum properties to process information more efficiently, establishing QML as a promising new paradigm \cite{Khan:2020}.
Basically, quantum computing offers distinct advantages over classical computing in addressing complex optimization and machine learning tasks \cite{RiveraRuiz:2022}. By exploiting the parallelism enabled by quantum effects such as entanglement and superposition, quantum algorithms can process vast datasets simultaneously, potentially providing exponential speedup for specific computations. This capability can significantly enhance the performance of machine learning models by efficiently exploring large solution spaces.
For instance, one can train the quantum circuit to achieve maximal separation between the data clusters in the Hilbert space, paving the way for the development of robust quantum classifiers. Quantum computing's ability to handle high-dimensional data and capture complex patterns makes it particularly suitable for applications such as traffic forecasting, where intricate relationships within the data must be captured.
To accelerate progress in QML, exploring heuristic algorithms, such as those employed in this study, is essential. Although these algorithms currently lack formal theoretical backing, they have demonstrated effectiveness in certain problem domains through cross-disciplinary insights and domain expertise.

Additionally, novel techniques such as data re-uploading \cite{EasomMccaldin:2021}, which involves encoding classical data into quantum circuits multiple times, can significantly enhance the expressivity of quantum models without drastically increasing computational demands. This technique holds great promise for the future capabilities of QML.
Data re-uploading is achieved by catenating repeating units in sequence. Single-qubit rotations applied multiple times within the circuit generate the necessary non-linearity to construct a functional neural network. A quantum circuit can then be organized as a series of data re-uploading and single-qubit processing units. Moreover, it has been demonstrated that a single qubit data-reuploading circuit can be utilised both as being a universal quantum classifier\cite{perez2020data} and being a universal approximant\cite{perez2021one}. Several recent studies indicate that data re-uploading can positively impact the performance and trainability of quantum models.

\begin{figure}[htbp]
\centerline{\includegraphics[width=0.5\textwidth]{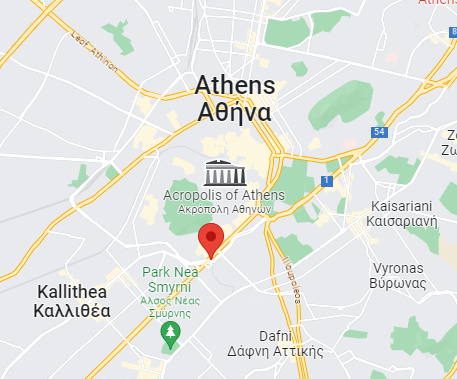}}
\caption{Location of the loop detector used in this study, relative to the center of Athens, Greece.}
\label{fig1}
\end{figure}

\bigskip
\noindent
In this paper, we propose the application of Quantum Machine Learning (QML), including data re-uploading, to the challenge of traffic forecasting. Our primary goal is to investigate whether a \textit{hybrid} quantum-classical neural network architecture can match or outperform purely \textit{classical} neural networks in predicting traffic flow.
As a case study, we utilize high-resolution traffic data from the city of Athens, Greece (Figure \ref{fig1}).

The core of our approach involves two distinct scenarios. In Scenario A, we focus on fully connected neural networks (NNs), replacing a classical fully connected layer with a quantum layer. This scenario allows us to investigate whether quantum layers can encode information more efficiently and capture the underlying traffic flow patterns better than classical counterparts.

 Scenario B explores recurrent networks, where we integrate quantum layers with data re-uploading. This technique mimics the recursive structure of LSTMs by re-uploading input data multiple times into the quantum circuit. This allows us to assess the capability of quantum layers to model time series data by capturing both short- and long-term dependencies within the traffic data.

We compare the performance of these hybrid models with their fully classical
counterparts in terms of forecasting accuracy and computational efficiency. The models are trained and tested on real-world traffic data with a time resolution of 1.5 minutes, covering 40 days of traffic flow collected from one of Athens' busiest roads.
To ensure a rigorous evaluation, we apply a 5-fold gap cross-validation protocol, which helps prevent temporal leakage between the training and testing sets and enhances the generalizability of our findings.

Our results demonstrate that while classical models remain highly efficient in terms of computational complexity, hybrid quantum-classical models show promise in improving forecasting accuracy, especially as the complexity of the quantum layer increases. Furthermore, we provide evidence that using quantum variational circuits in combination with data re-uploading allows the model to better capture the complex patterns and temporal dependencies in the traffic data, potentially offering a novel way forward for traffic forecasting in ITS.

\bigskip
\noindent
The contribution of this paper is, therefore, twofold:
\begin{itemize}

   \item We present one of the first applications of QML to traffic forecasting, demonstrating the feasibility and potential of quantum-enhanced models for real-world forecasting tasks.
   Moreover, to our knowledge, this is the \textit{first study} that successfully employs data re-uploading to a traffic prediction scenario.
  \item  We offer a detailed comparison between classic and quantum approaches, highlighting the strengths and weaknesses of each and providing insights into the conditions under which quantum models may offer a tangible advantage.

\end{itemize}

The paper is organized as follows: Section \ref{Related Work} provides a comprehensive review of existing research on DL methods for traffic forecasting and explores the emerging role of quantum computing in transportation. Section \ref{methods} details our model architectures, data sources, and the hybrid quantum-classical approach we employed. In Section \ref{results}, we present our findings, including performance metrics and insights into the convergence behavior of the models. Finally, Section \ref{Discussion} discusses the implications of our results and outlines future research directions, while Section \ref{conclusion} concludes the paper with a summary of key contributions.

\section{Related Work}\label{Related Work}

\subsection{Classical Deep Neural Networks for Traffic Forecasting}

Over the past few decades, significant advancements in telecommunication technology and computing systems have enabled researchers and practitioners in the field of traffic prediction to increasingly focus on DL. This natural progression has been driven by the growing availability of vast amounts of relevant data, further facilitating the adoption and utilization of DL techniques in this domain (e.g., \cite{Zhao:2017, Bapaume:2021}). One of the primary advantages attributed to DL methods is their notable superiority in prediction accuracy compared to traditional Statistical and Machine Learning approaches \cite{Wang:2019}.

Currently, in the realm of traffic conditions forecasting, Recurrent Neural Networks (RNNs) and, more specifically, LSTM networks, have emerged as the most widely adopted models. These models are not only popular but are often combined with other architectures to achieve remarkably precise predictions (e.g., \cite{Fafoutellis:2020, Ranjan:2020}). A drawback associated with these models is that, when dealing with long sequences, their capacity to retain information from distant-past timesteps may diminish, a phenomenon known as the ``vanishing gradient issue''\cite{Yin:2021}. Furthermore, while the RNN module and its variations excel in handling time-series problems, they are limited in their ability to capture spatial relationships within traffic data (e.g., \cite{Boukerche:2020}).

Conversely, CNNs, primarily employed in image recognition and computer vision tasks, are harnessed in traffic forecasting to effectively leverage spatial relationships (e.g., \cite{Yin:2021}). In implementing CNNs, the road network is typically represented as a 2-dimensional grid, similarly to an image. However, since this representation is effectively static, a common approach is to combine CNNs with RNNs. This combination allows for the exploitation of consecutive images, enabling a better understanding and modeling of temporal dynamics in traffic forecasting (e.g., \cite{Ma:2017, Dai:2019}).

Recently, Graph Convolutional Neural Networks (GCNNs) have emerged as an alternative to CNNs. Unlike CNNs, which operate in the Euclidean domain and represent the road network as an image, GCNNs extend the convolution operation to accommodate more general graph-structured data, making them more suitable for effectively representing a road network \cite{Jiang:2021}. The model’s input consists of the graph-structured data's adjacency matrix, which not only reflects the nodes' connectivity but may also capture statistical correlations. Additionally, a set of features is provided for each node, such as measured traffic flow, speed, and more. GCNNs have proven to be the current state-of-the-art in traffic prediction and have been prominently employed in several recent and noteworthy research endeavors (e.g., \cite{Leiser:2021, Cui:2020, Yu:2021}).

Despite the high accuracy of Deep Neural Networks compared to previous approaches, their applicability still faces several challenges, which are outlined below.

\begin{itemize}
    \item Significant data requirements $-$ Deep Neural Network models demand a substantial volume of data encompassing various traffic conditions for effective training and convergence. Moreover, the data must be extensive and diverse enough to ensure that the model's generalization capability remains uncompromised \cite{Yin:2021}.
    \item Extended training time $-$ The complex architecture of Deep Neural Networks, with numerous layers and numerous hyperparameters, contributes to longer training times compared to traditional statistical and simpler Machine Learning models. Additionally, updating and retraining the models' parameters when new data becomes available is a time-consuming and resource-intensive process (e.g., \cite{Boukerche:2020}).
    \item Hyperparameter selection challenges $-$ Determining the appropriate number of hidden layers and neurons in each hidden layer in a Neural Network often relies on experience or a trial-and-error process. Having too many neurons per hidden layer leads to an overfitting-prone network and prolonged computational times. Conversely, using too few neurons can compromise prediction accuracy, especially when dealing with a substantial volume of input data.
    The absence of a definitive solution for determining the optimal architecture remains a persistent issue.    
\end{itemize}

These factors explain why research on DL methodologies remains an extremely active field and motivates studies like the current one.
Striking a balance between model complexity, computational resources, and time is crucial when deploying Neural Networks in real-world conditions, as these factors can influence the model's accuracy. The aforementioned limitations apply not only to Neural Networks in general but also hold significance in the context of traffic data. Collecting adequate data from the entire road network over an extended period can be challenging. Moreover, the lack of interpretability can restrict the practical applicability of prediction models, as previously noted.

\subsection{Quantum computing in transportation}

Quantum computing has the potential to revolutionize transportation by solving computational challenges intractable for classical computers. Over the past few years, quantum computing has garnered significant attention for its potential applications in transportation optimization. Several studies have explored this promising technology for various transportation problems. For instance, the authors in \cite{Wang:2021} examined the future potential of quantum computing in ITSs, highlighting its ability to tackle complex computational problems. Bentley et al.\ \cite{Bentley:2022} demonstrated the application of quantum computing for optimizing transport routes, showcasing improvements in computational efficiency and solution quality.
Similarly, the study in \cite{Dixit:2023} applied quantum computing to transport network design problems, illustrating its advantages in handling large-scale network optimization tasks. The study by \cite{DixitRouting:2023} focused on using quantum computing to solve scenario-based, stochastic, time-dependent, shortest-path routing problems, further proving its efficacy in dynamic and uncertain environments. Additionally, Yarkoni et al.\, in \cite{Yarkoni:2020}, introduced the ``Quantum Shuttle'' system, which utilized quantum computing for real-time traffic navigation during large events, marking a significant milestone as the first commercial application of quantum computing in traffic management.


\subsection{Quantum Machine Learning}

One of the most plausible candidates for exploiting the practical advantages of quantum computing in the NISQ era is QML \cite{Brooks:2019}. QML offers a wide range of applications, such as utilizing data-driven approaches to discover quantum algorithms, optimizing quantum experiments, processing classical or quantum information using Quantum Neural Networks (QNNs), and even developing quantum-inspired classical Machine Learning protocols \cite{Cerezo:2022}. Various QML techniques, including QNNs with parameterized quantum circuits and measurements, hybrid quantum-classical schemes, and quantum heuristic algorithms, have been proposed to tackle these tasks \cite{Biamonte:2017, Cai:2022, Haug:2021, Mari:2020}.

 In our previous works (see \cite{Schetakis:2022} and \cite{SchetakisCredit:2022}), we provided a comprehensive discussion on the role of quantum layers in hybrid neural networks and their effect on the learning process. We addressed how QNNs can improve a task's performance by utilizing the inherent advantages of quantum mechanics, such as the ability to process information in more complex ways than classical networks. By increasing the complexity of these quantum layers, the networks could better capture and model intricate patterns in the data, potentially leading to enhanced predictive accuracy.

However, despite the rapid progress in the field, numerous open and challenging tasks remain to be addressed. These include efficient encoding data schemes for quantum processing, improving quantum models, refining training methodologies, enhancing generalization capabilities, mitigating the impact of quantum noise, and more. The present research aims to tackle several ongoing research issues in QML. Drawing on the practical advantages of quantum computing during the NISQ era, we seek to explore innovative solutions for data encoding, quantum model optimization, robust training techniques, and other related challenges. By addressing these issues, we seek to advance the capabilities of QML and unlock its full potential for real-world applications in the NISQ era.

\section{Methodology}\label{methods}

\subsection{Data Description}

\begin{figure*}[htbp]
\centering
  \includegraphics[width=0.7\textwidth]{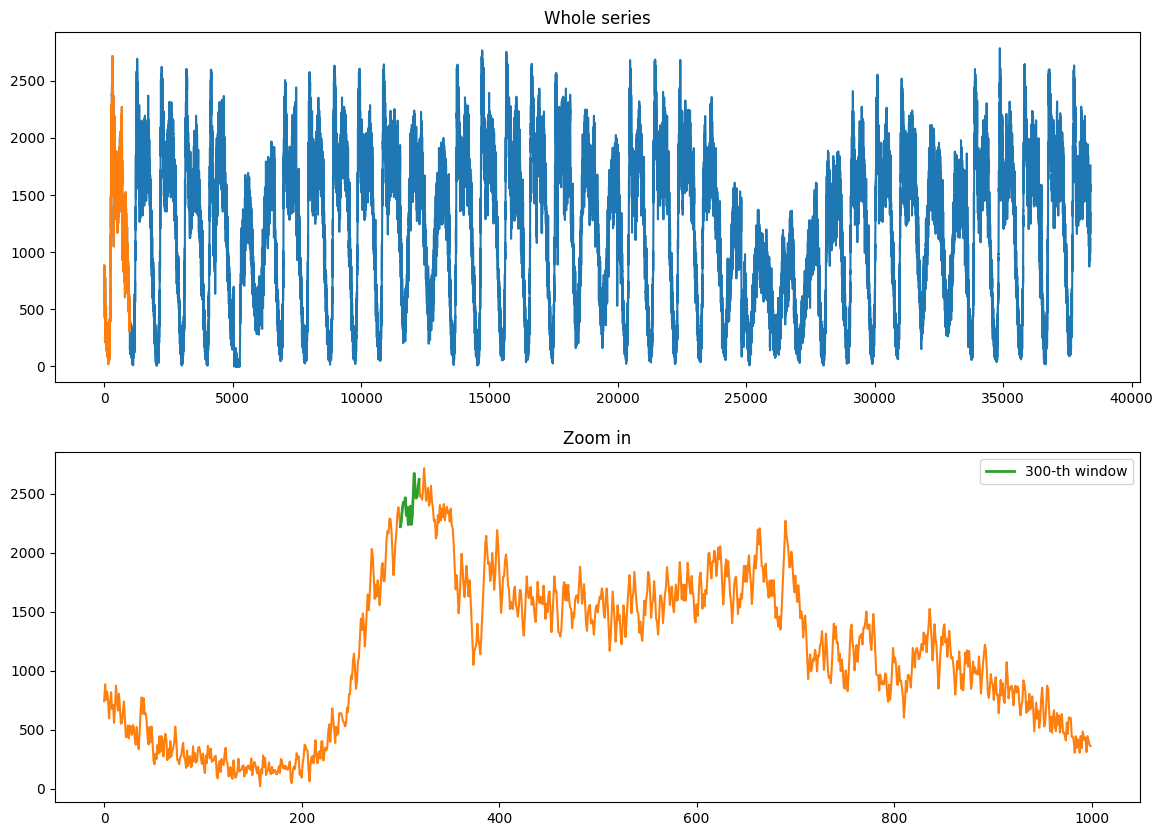}
  \caption{
  Full extent (\textit{top}) and detailed view (\textit{bottom}) of the time series used in the study.
  The $x$-axis displays the datapoint index, while the $y$-axis shows the traffic volume in units of vehicles/hour.
  }
  \label{figure:dataset}
\end{figure*}

Our traffic forecasting task was based on traffic flow data representing the number of vehicles passing through a loop detector installed on Syggrou Avenue, situated in the heart of Athens, Greece. This road, which leads toward the city center, is one of the busiest in the metropolis, with three lanes accommodating over 2,000 vehicles per hour during peak times. This location was selected for this study specifically due to its high traffic volume, which makes it an ideal subject for analysis. The exact position of the loop detector is shown in Figure \ref{fig1}.

The data were sourced from a database jointly developed by the Greek Government and the Region of Attica, which compiles information from nearly 400 loop detectors across the region. This dataset was provided for academic research and relevant studies. For this study, we analyzed 40 days of traffic flow data collected during March and April 2023, with a time resolution of 1.5 minutes, corresponding to 40 measurements per hour. Traffic flow (or traffic volume) represents the number of vehicles passing through a specific point on the road network within a given time frame, typically expressed in vehicles per hour. This metric varies between approximately 2,000 vehicles/hour for large highways and arterials, to as low as 500 vehicles/hour for urban road sections, and even less during congested periods. 

The complete time series is illustrated in Figure \ref{figure:dataset}, where the top panel presents the full extent of the data. The $x$-axis represents a timestamp index, while the $y$-axis indicates vehicle flow in units of vehicles per hour. A closer view on a specific time series segment is shown in the bottom panel of the same figure, with the green section highlighting the size of a ``windowed'' unit used during the training phase of the neural network model (see Section \ref{Training}).

\subsection{Model Architectures}
\label{Model Architectures}

The main objective of this study is to compare classical against hybrid quantum-classic Neural Network (NN) approaches in a traffic forecasting task. 
The underlying goal is to assess the capacity and efficiency of quantum layers in NNs and examine whether such quantum configurations offer tangible advantages over their classical counterparts.

Quantum layers resemble classic ones in their high-level usage but fundamentally differ in their operational details.  
At the core of the quantum layer lies a Quantum Variational Circuit (QVC), which is central to the hybrid quantum-classical NN architecture.
The QVC is designed to leverage the principles of quantum mechanics and is usually composed by the chaining of three key elements (e.g., \cite{Bharti:2022,CerezoVQA:2021}):

\begin{enumerate}
    \item \textit{Data Embedding ---}
    The data embedding stage is responsible for encoding the classical data into quantum states.
    Regardless of whether the classical data serves as the network input or comes from the preceding classical layer (referred to as the ``feeding layer''), the embedding stage encodes this information into qubits.
    One way to do this is via angle rotation encoding, a technique where classical values are mapped to the rotational angles of qubits.
    These rotations prepare the quantum state that subsequent quantum operations will process.
    The role of this encoding step is crucial, as it directly impacts the ability of the quantum circuit to capture the nuances of the data.
    \item \textit{Entangling stage ---}
    Following the angle embedding, the qubits undergo a series of operations in the entangling layer.
    As an example, in our models (see Sections \ref{Scenario A} and \ref{Scenario B}), we will use rotational gates and Controlled-NOT (CNOT), which are building blocks for creating quantum entanglement.
    The CNOT gates generate entanglement between pairs of qubits, allowing the quantum layer to capture complex interdependencies within the data.
    The rotational gates, which have trainable parameters, adjust the quantum states further, based on the encoded data.
    The entangling process is a key aspect of quantum computing that enables the model to explore a much larger solution space than classical methods could, potentially leading to better generalization and more accurate predictions.
    \item \textit{Measurement Stage ---}
    The final component of the QVC is the measurement stage, where the quantum states are measured to extract classical information from the quantum layer. Measurement collapses the quantum states into classical bits, which are then passed to the next layer in the NN. This step bridges the quantum and classical parts of the hybrid model, allowing the information processed by the quantum layer to inform the subsequent operations in the classical layers. The measurement quality directly affects the quality of the information the classical network receives, making it a critical part of the quantum-classical interface.
\end{enumerate}

These operational peculiarities of the quantum layers suggest that conducting a \textit{classical vs.\ hybrid} model comparison is a challenging task.
In practice, this difficulty boils down to the issue that the classical equivalent to a qubit -- the fundamental computational unit in quantum architectures -- is not easily defined, if it can be defined at all.
In fact, the two computational techniques are intrinsically different in both the way they store \textit{and} process the information.
Therefore, when performing such comparisons, one has to decide on which concept the classic--hybrid equivalence shall be based on.
In this study, we opted to contrast the capabilities of classical and quantum networks whose complexities have been deemed comparable under the principles of \textit{A}) size of encoded information, and \textit{B}) number of recursive iterations.
Namely, to explore these two scenarios, we designed two distinct experiments in which we replaced a layer of completely classic NNs with their quantum equivalents.
In the first case, we focus on NNs based on fully connected layers, while in the second, we explore recursive NNs.

\subsubsection{Scenario A: Layer replacement based on equal amount of encoded information}
\label{Scenario A}

\begin{figure*}[htbp]
    \centering
    \includegraphics[width=\textwidth]{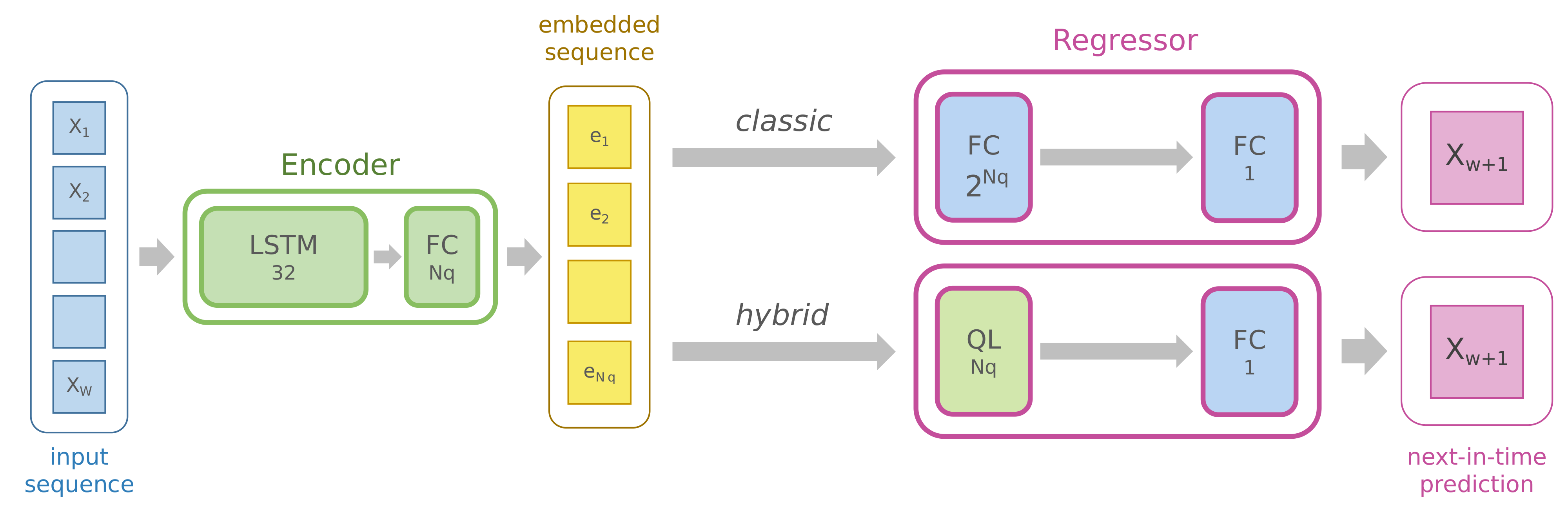}
    \caption{
    Schematic representation of the fully connected architecture used in the scenario described in Section \ref{Scenario A}.
    The left side of the image shows the encoding provided by the autoencoder trained as depicted in Figure \ref{figure:autoencoder}.
    The right side represents the part of the NN that acts as a regressor.
    In Section \ref{Scenario A}, we compare two approaches: a fully \textit{classic} one (\textit{top right}), and a \textit{hybrid} one (\textit{bottom right}): the difference between the two lies in the first fully connected layer.
    In either case, the output of the NN is a single value representing the prediction at the timestep immediately following the input sequence.
    }
    \label{figure:FC}
\end{figure*}

\begin{figure}[htbp]
    \centering
    \includegraphics[width=0.8\textwidth]{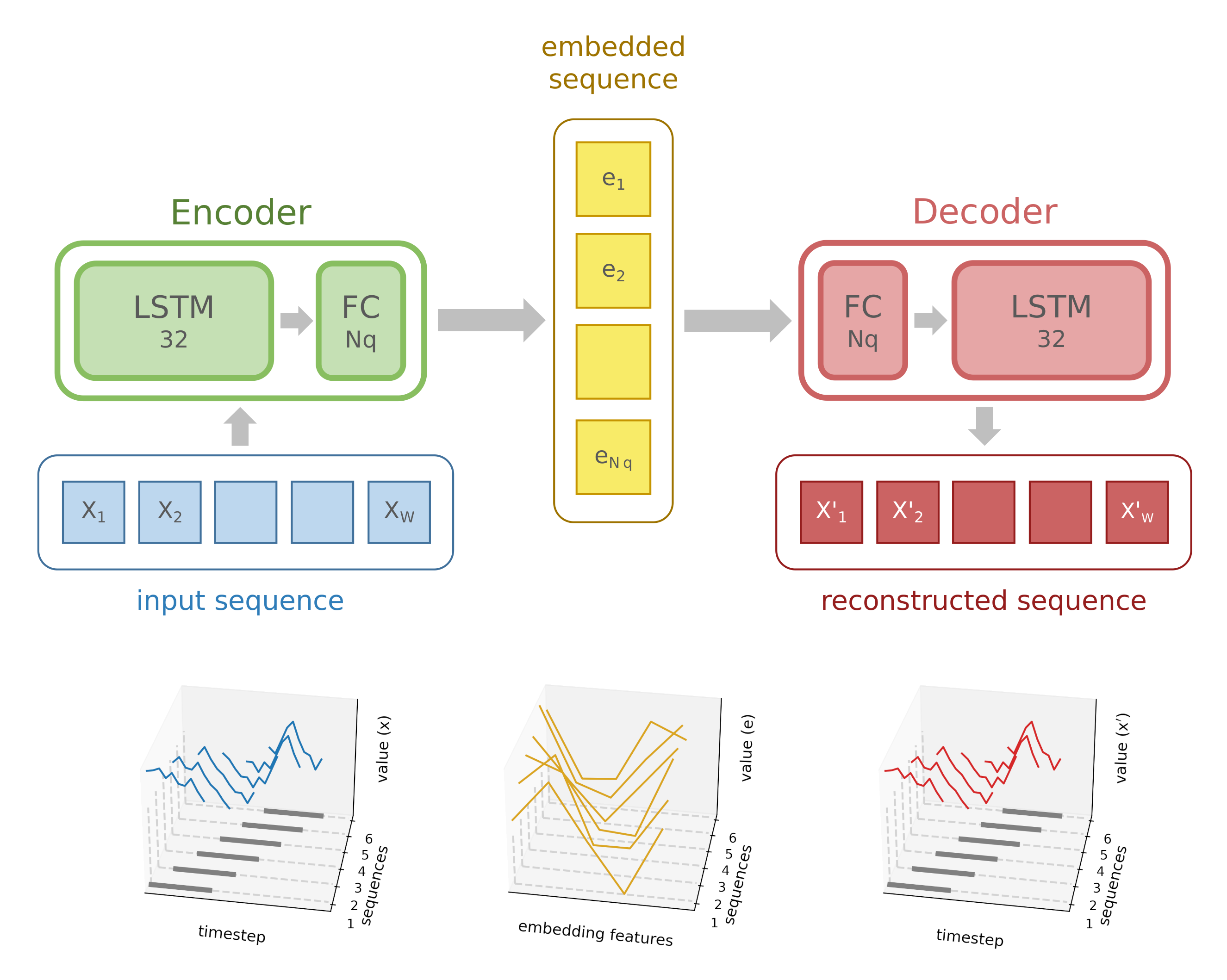}
    \caption{
    The architecture of the autoencoder is used as a preprocessing step for the NNs in Figures \ref{figure:FC} and \ref{figure:RNN}.
    The encoder is constituted by an LSTM cell composed of 32 units and an FC layer, which shrinks the embedding space to the desired $N_q$ features; the same structure is mirrored in the decoder. 
    The autoencoder is trained in a standard way, i.e., by matching the input sequence with the reconstructed one via minimization of the mean squared error.
    }
    \label{figure:autoencoder}
\end{figure}

\smallskip
In this scenario, we consider an NN in which the burden of handling the regression task falls primarily on the Fully Connected (FC) layers, which serve as the core computational components responsible for processing and refining the input data.
For a detailed overview of this scenario’s architectures, refer to Figure \ref{figure:FC}, which illustrates the structure and interplay of these FC layers within the broader network.

In this context, we aim to explore the effects of substituting a single, conventional, feed-forward FC layer of an NN \textit{with} a quantum layer designed to withhold an equivalent amount of information.
The underlying premise is that a quantum system, through the use of qubits, has the potential to exploit higher-dimensional embedding spaces.
Specifically, $N_q$ qubits create an embedding space of dimensionality $2^{N_q}$, allowing us to theoretically replace a classical layer comprising $N_q$ neurons with a quantum layer consisting of approximately $\sqrt{N_q}$ qubits.
However, for the practical implementation, we reverse this approach (without loss of generality): we begin by constructing a quantum layer with $N_q$ qubits and subsequently compare its performance against a classical layer containing $2^{N_q}$ neurons.

Exploring the performance of FC layers in a time series example is challenging because FC layers fall short when it comes to handling temporal dependencies.
This makes them less suitable for tasks where the sequence and timing of data play a crucial role, as they do not inherently account for the dynamic patterns and interactions that unfold over time.
To fight this limitation, we can introduce a form of preprocessing that precedes the FC layer and is capable of capturing the temporal dependencies.
Specifically, we adopted an autoencoder architecture based on LSTM cells.
After training, the encoder is able to process input sequences, reduce their dimensionality, and, crucially, account for time-dependent relationships.
We stress that, while dimensionality reduction is not strictly necessary in this stage -- its main objective being to capture the temporal properties -- it becomes a beneficial side effect, as it enhances computational efficiency by reducing the overall complexity of the input, thereby speeding up subsequent calculations.

The autoencoder component is shown in Figure \ref{figure:autoencoder}, while the complete pipeline architecture is displayed in Figure \ref{figure:FC}, and works as follows.
Every ``input sequence'' ($x_1$, $x_2$..., $x_w$), where $w$ is the window size (in our case, 20 timesteps), is first parsed through the pre-trained encoder to produce a compressed ``embedded sequence'' ($e_1$, $e_2$..., $e_{N_q}$) of size ${N_q}$.
The embedded sequence is then passed to the regressor, which, as described above, can be the classical regressor, in which the first layer is an FC layer composed of 2$^N_q$ neurons, or the hybrid regressor, in which the first layer is a quantum layer composed of $N_q$ qubits.
The final output is, either way, a single classic neuron with a linear activation function, as the NN is designed to provide for the next-in-time prediction.

\subsubsection{Scenario B: Layer replacement based on equal amount of recursions}
\label{Scenario B}

\begin{figure*}[htbp]
    \centering
    \includegraphics[width=\textwidth]{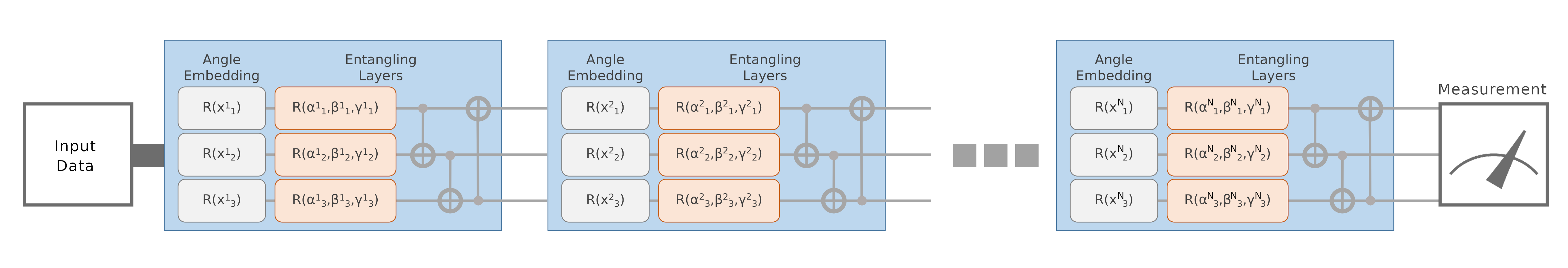}
    \caption{
    Example of an unfolded data re-upload scheme with $N$ re-upload blocks (blue boxes) with the same components as the ones adopted in this work.
    In this specific depiction, the classic input data have dimensionality 3, and the circuit is composed of 3 qubits.
    In each block, the grey boxes represent the rotations used to embed the classical data, and two entangling layers follow them.
    The first such layer is composed of rotational gates (orange boxes), each characterized by three tunable parameters ($\alpha$, $\beta$, and $\gamma$), while the second layer is composed of CNOT gates.
    At the output of the circuit, a measurement converts the signal back to classical data (e.g., in our work, we applied a Pauli-Z measurement to observe the state of the qubits along the Z-axis in the computational basis).
    In this representation, variables and parameters are indexed as $\langle . \rangle^{n}_{q}$, where $n$ is the block index, and $q$ the qubit index.
    In general, multiple entangling layers may be chained inside a single block to increase its complexity.
    }
    \label{figure:Reupload}
\end{figure*}

\begin{figure*}[htbp]
    \centering
    \includegraphics[width=\textwidth]{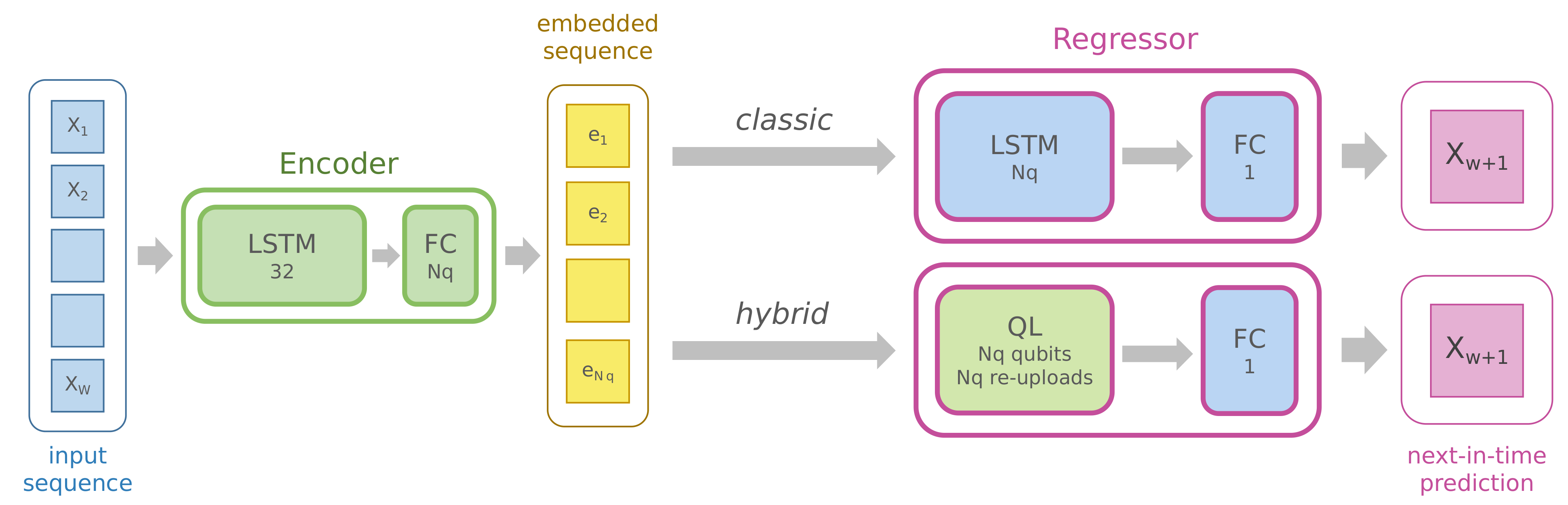}
    \caption{
    Schematic representation of the recursive architecture used in the scenario described in Section \ref{Scenario B}.
    The left side of the image shows the encoding provided by the autoencoder trained as depicted in Figure \ref{figure:autoencoder}.
    The right side represents the part of the NN which acts as a regressor.
    In Section \ref{Scenario B}, we compare two approaches: a fully \textit{classic} one (\textit{top right}), and a \textit{hybrid} one (\textit{bottom right}): the difference between the two lies in the first layer.
    In the \textit{classic} approach, the first layer is an LSTM cell composed of $N_q$ units, while in the \textit{hybrid} case, it is a quantum layer with $N_q$ re-uploads and $N_q$ qubits.
    In either case, the output of the NN is a single value representing the prediction at the timestep immediately following the input sequence.
    }
    \label{figure:RNN}
\end{figure*}

\smallskip
In this scenario, we attempt to go beyond fully connected layer architectures and use recursive layers for the regression task.
Specifically, we design the regressor, which contains an LSTM cell, and compare it against a regressor in which this element has been replaced by a quantum layer characterized by multiple data \textit{re-uploading}.

Data re-upload \cite{EasomMccaldin:2021} is a technique initially designed to enhance the learning capacity of the model without increasing circuit depth or qubit count. 
It works by encoding the same classical data into the quantum circuit multiple times, as depicted in the example of Figure \ref{figure:Reupload} and described by the following pseudo-algorithm:

\begin{algorithm}[H]
\caption{Data Re-upload Pseudo-Algorithm}
\begin{algorithmic}
\STATE Initialize quantum state $\vert \psi \rangle$
\FOR{each re-upload $r$}
    \STATE 1. Encode classical data $x$ using parameterized gates
    \newline 
    \hspace*{0.2cm} \textit{$R_x(x, \theta_r), R_y(x, \theta_r)$}
    \STATE 2. Apply quantum operations
    \newline 
    \hspace*{0.2cm} \textit{(e.g., rotations, entangling gates)}
\ENDFOR
\STATE Measure final quantum state
\end{algorithmic}
\end{algorithm}

Intuitively, since each re-upload uses different parameters for the tunable gates ($\theta_r$) at each iteration, data re-uploading allows to explore more complex patterns.
Not surprisingly, since its first appearance, this technique acquired immediate success due to its potential to improve performance, and it has been applied in diverse classification and regression tasks (e.g., \cite{Schuld:2021} and references therein).

However, data re-upload has been exploited so far primarily for its potential to explain complex patterns; in this work, we intend to shift the attention to another property of this methodology.
In fact, note that the same classical data are iteratively layered onto an already evolved quantum state (modified by the ``quantum operations'' in between re-uploads).
Therefore, data re-uploading does not \textit{erase} the previous quantum state, but rather \textit{adds} to it.
In this sense, while the quantum system retains the previously acquired quantum information, the re-uploading allows further manipulation of the state.
This technique, hence, works in a similar fashion to recursive memory cells in classic NNs, such as LSTMs.

The complete pipeline architecture used in this experiment is displayed in Figure \ref{figure:RNN}.
As in Scenario A, the input sequence is first parsed through the pre-trained encoder to produce a compressed embedded sequence, which is, in turn, passed to the regressor.
In the classic regressor, the first layer is an LSTM layer composed of $N_q$ units; notice that, since the embedded sequence is also $N_q$ timesteps long, this LSTM will recurse $N_q$ times.
Similarly, the first layer of the hybrid regressor is a quantum layer acting on $N_q$ qubits and performing $N_q$ re-uploads (re-uploading ``blocks'')\footnote{
This limitation, i.e., that of picking $N_q$ both for the number of qubits \textit{and} the number of re-uploads, is dictated by the way the data re-uploading in hybrid networks is implemented in Pennylane -- see Section \ref{Further insights on the quantum layers}.
}.
Notice that, in our experiments, each re-uploading block consists of an angle embedding layer followed by a single entangling layer, although, in principle, one could add multiple entangling layers in each block.

We emphasize that, by configuring the networks in this manner, we effectively propose an innovative solution to the challenge of comparing quantum and classical recursive networks.
In fact, the LSTM and quantum elements that we designed are equivalent in the sense that both take as input $N_q$-long sequences, produce $N_q$-long sequences as outputs, and recurse $N_q$ times.
The final output is either way a single classic neuron with a linear activation function, as the NN is designed to provide for the next-in-time prediction.

\subsubsection{Further insights on the quantum layers}
\label{Further insights on the quantum layers}

\smallskip
Simulating quantum layers on classical hardware is a computationally intensive task, especially as the number of qubits increases. Simulating more than 10 qubits can quickly become unfeasible on standard personal computers, making it necessary to strike a balance between the model’s complexity and the computational resources required.
For this reason, we kept the quantum layers relatively small in terms of qubits ($N_q \le 14$), acknowledging that this choice might come at the expense of the model’s potential performance.
However, as demonstrated in our results (Section \ref{Performance Scores}), our model architecture achieved extremely high-performance scores, primarily due to the abundance of training data available.

The quantum layers were implemented using Pennylane\footnote{
    \href{https://pennylane.ai}{\texttt{https://pennylane.ai}}
}, a Python-based tool specifically designed for quantum machine learning and optimizing hybrid quantum-classical computations \cite{Schetakis:2022, Bergholm:2018}.
Pennylane provides the necessary infrastructure to simulate quantum circuits on classical computers, making it possible to explore the potential of quantum computing in practical applications, even without access to physical quantum hardware.

Note that, in the remainder of this manuscript, the classic--hybrid architecture pairs will be identified by the number of qubits in the \textit{hybrid} neural network (e.g., `Q2’ indicates 2 qubits), with the number of neurons in the corresponding layer of the classic NN being derived as described in Sections \ref{Scenario A} and \ref{Scenario B}. 

\subsection{Training}
\label{Training}

The NNs described earlier are designed to predict the value of the series at a specific timestep $t + 1$ based on the values from the previous $w$ timesteps ($x_{t-w}$, $x_{t-(w-1)}$, \ldots, $x_{t}$).
To achieve this, the dataset is divided into ``windows'', which are small sequences of $w$ data points, with each window serving as a predictive variable array.
The next data point in time, immediately following each window, is the target variable the model attempts to predict.

The training process involves feeding these windows and their corresponding target values into the network in mini-batches.
This approach helps stabilize and accelerate the learning process, especially when dealing with large datasets.
An Adam optimizer with a learning rate of 0.0005 has been adopted to minimize the Mean Squared Error (MSE) -- the loss function -- helping the network learn the mapping between input sequences and their corresponding targets.
The prediction stage follows the same rationale, with the $t+1$ test data point being predicted given the previous $w$ values.

\subsection{Performance Evaluation}
\label{Performance Evaluation}

\begin{figure}[htbp]
    \centering
    \includegraphics[width=\textwidth]{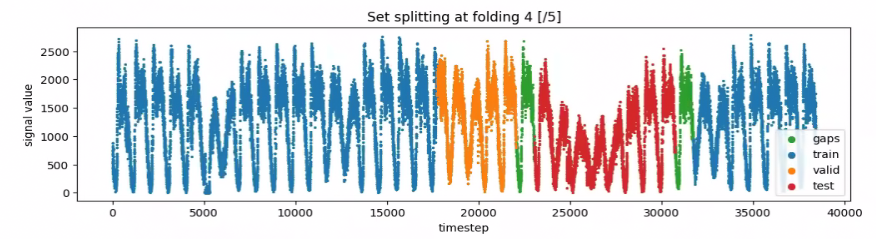}
    \includegraphics[width=\textwidth]{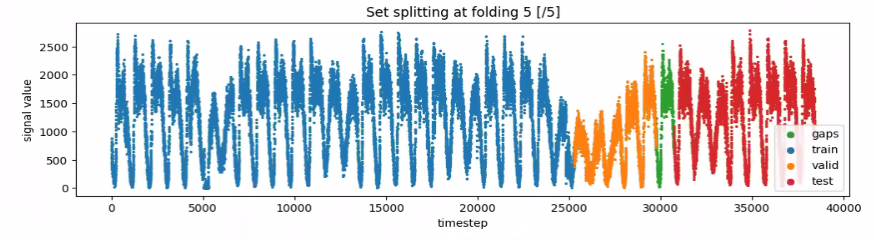}
    \caption{
    Two iterations of the gap-cross-validation protocol were employed in this work to assess the performance uncertainties. The train, validation, and test folds are represented in blue, orange, and red, respectively. The green slices show the gaps excluded from the analysis at each folding. The validation set is forced always to precede the test set, while the training set can be either earlier or later than the test set. The test set is always a unique block, but the other sets can wrap around. The gap is not necessary if the test fold appears at the edge of the series (e.g., bottom panel).
    }
    \label{figure:CV}
\end{figure}

Regarding the testing, we strived to obtain an unbiased estimate of the uncertainty about the results in order to quantify the difference between the NNs.
In time series forecasting modeling, it is common to segregate a test set that comes temporally after the training set.
This approach is most diffused because of the implicit assumption that time series data points are timely correlated and potentially causally correlated.
It follows that the value of the series at a given time $t$ is strongly associated with that at time $t-1$.

However, while it is not controversial to segregate a single hold-out test (at the end of the series) for an individual assessment, placing an uncertainty on this quantity is not straightforward.
In fact, this requires performing multiple train/test splits -- hence raising the question: where should the different test sets be selected along the time series?
In tabular data multiple testing is readily solved by using cross-validation (CV), in which the data ensemble is split in multiple folds, and one of the folds plays the role of test set (while the rest play the training), until all permutations are covered.
However, because of the aforementioned time-dependency issues, the same approach is not deemed applicable right away to time series, and, in general, there seems to be an open debate on what the best solution might be (see, e.g., \cite{Bergmeir:2018} and references therein).

In this study, we opted for a 5-fold gap-cross-validation technique \cite{Racine:2000} via the GapKFold\footnote{
\href{https://github.com/WenjieZ/TSCV}{\texttt{https://github.com/WenjieZ/TSCV}}
}
Python implementation \cite{Zheng:2023}.
This protocol is similar to the CV, with the significant difference that there are ``gaps'' (i.e., discarded data) on both sides of the test fold.
In other words, at each CV permutation, part of the data contiguously preceding and following the test fold are ignored.
Figure \ref{figure:CV} shows a graphical representation of our own variant of this protocol, in which,  at each folding, we also insert a validation set\footnote{
  The validation data are solely used to monitor the NN training.
}.
In summary, the core idea of this methodology is that discarding temporally adjoined data guarantees, to some extent, a break in the causal dependence between training and test, providing independent sets.
Notice that this technique uses -- as training data -- folds that may be located temporally after the test set.

The sequences (windows) used for the actual training (see Section \ref{Training} and Figure \ref{figure:dataset}, bottom panel) are generated after the train/validation/test splicing to guarantee that there is no data leakage across the sets.

\section{Results}\label{results}

\subsection{Training convergence}
\label{Training convergence}

\begin{figure*}[htbp]
    \centering
    \begin{minipage}{0.48\textwidth}
        \centering
        \includegraphics[width=\textwidth]{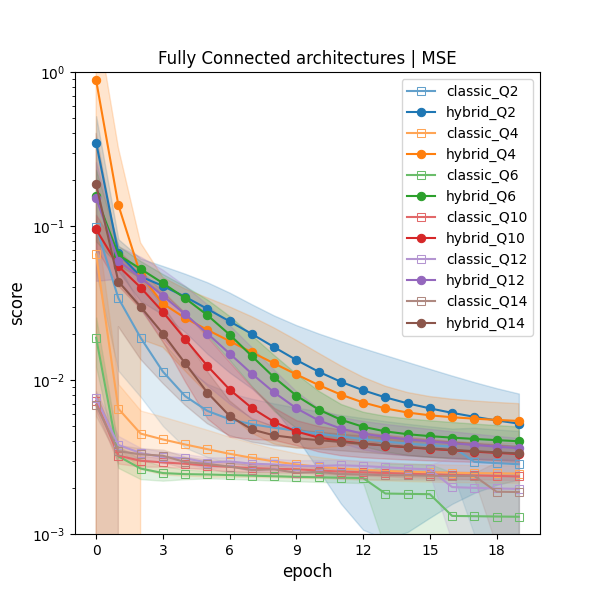}
    \end{minipage}
    \begin{minipage}{0.48\textwidth}
        \centering
        \includegraphics[width=\textwidth]{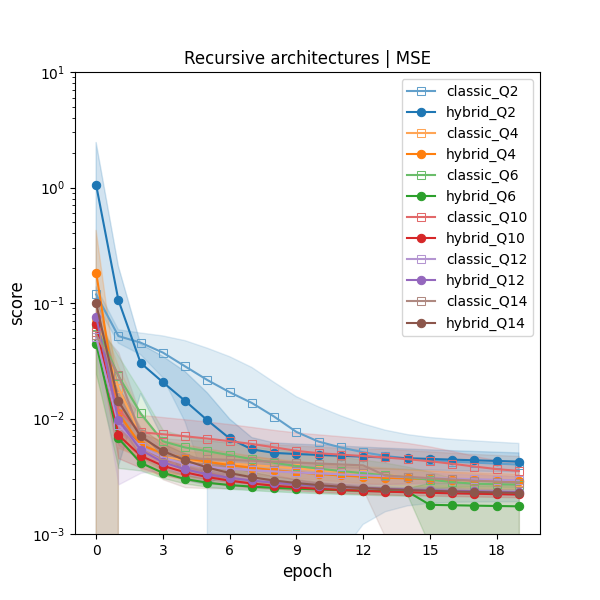}
    \end{minipage}
    \caption{
      Evolution of the models' loss (MSE) as measured on the training sets for the fully connected architectures (Scenario A; \textit{left}) and the recursive architectures (Scenario B; \textit{right}).
      The curves represent the mean value at each epoch, averaged over the 5-fold cross-validation.
      The shaded areas, corresponding to the color of each curve, indicate the 1-$\sigma$ standard deviation around the mean.
      Empty circles represent \textit{classic} architectures, while filled circles of the same color refer to the corresponding \textit{hybrid} ones.
      The scores refer to the normalized data values.      
     \label{figure:histories}  
    }
\end{figure*}

One of the key challenges of this study is to compare the computational speed of classic NNs as opposed to hybrid ones, i.e., examine which one converges faster.
Due to the significant difference in computational requirements -- given that quantum layers are being simulated on classical hardware -- a direct comparison of wall clock times would be unfair.
Instead, our study compares the speed of convergence in terms of training epochs, providing a more meaningful metric in this context.
In Figure \ref{figure:histories} we show the evolution of the training loss (MSE) across the epochs, for each model considered in this work.

Two main considerations emerge from this figure.
In the first place, we notice that both fully connected and recursive architectures manage to converge within the 20 epochs we experimented on, and moreover converge to very similar scores.
This relatively rapid convergence is due, more than to a complex network architecture, to the extremely large wealth of data available for our series.
We had about 40\,000 data points, which we split into windows of 20 data points each, ultimately resulting in approximately 800 batches (composed of 32 windows each) per epoch.
Regarding the similar score values, consider that this dataset exhibits an extremely regular pattern, and that the task is a next-in-time forecasting challenge, where the goal was to predict a single future data point per window. 
These factors greatly simplified the forecasting problem, thereby explaining the consistently high performance across all NNs.
However, note that while the curves of all recursive models (except for Q2) flatten at around epoch 5, the curves for the fully connected architectures present a larger spread.

This is connected with the second observation, i.e., that the hybrid models in fully connected architectures converge way slower than their corresponding classical ones. 
This trend is reversed in the recursive architectures where, although less significantly, the hybrid models converge faster than their counterparts, especially for smaller numbers of qubits.

\subsection{Performance Scores}
\label{Performance Scores}

\begin{figure*}[htbp]
    \centering
    \includegraphics[width=\textwidth]{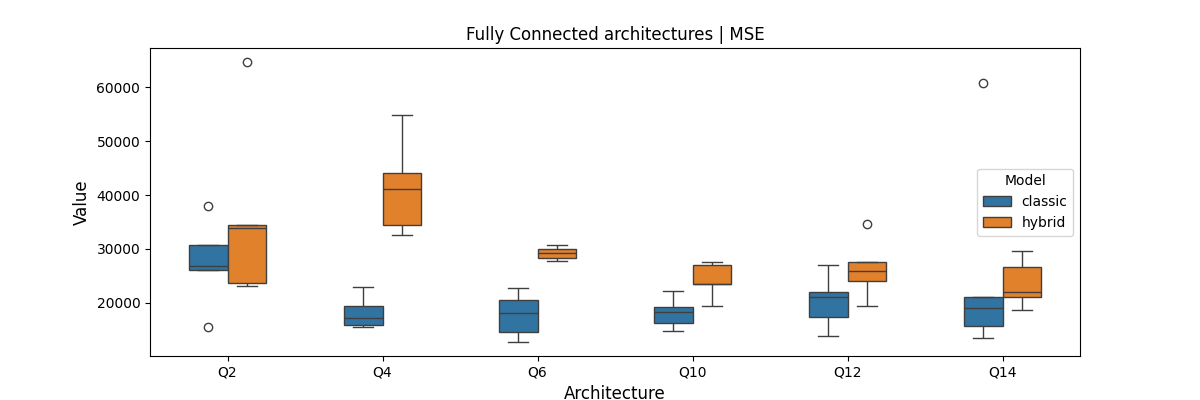}
    \includegraphics[width=\textwidth]{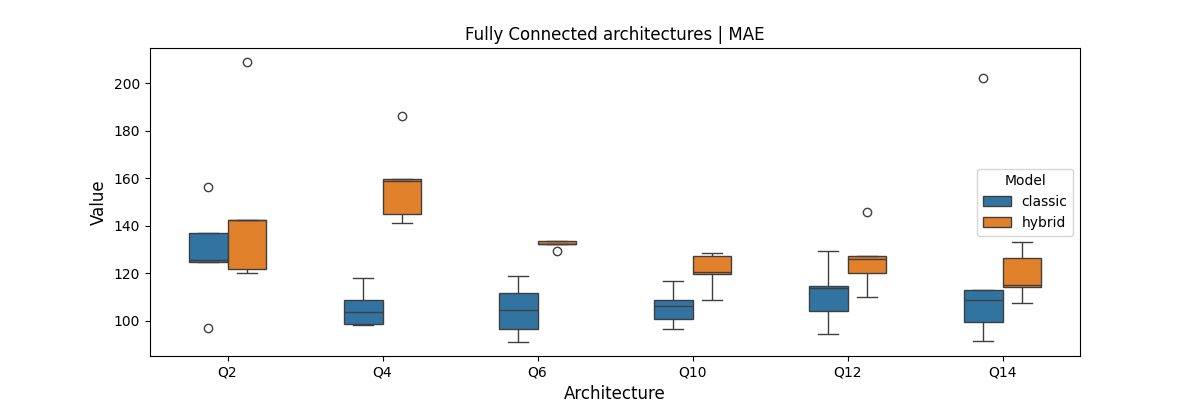}
    \includegraphics[width=\textwidth]{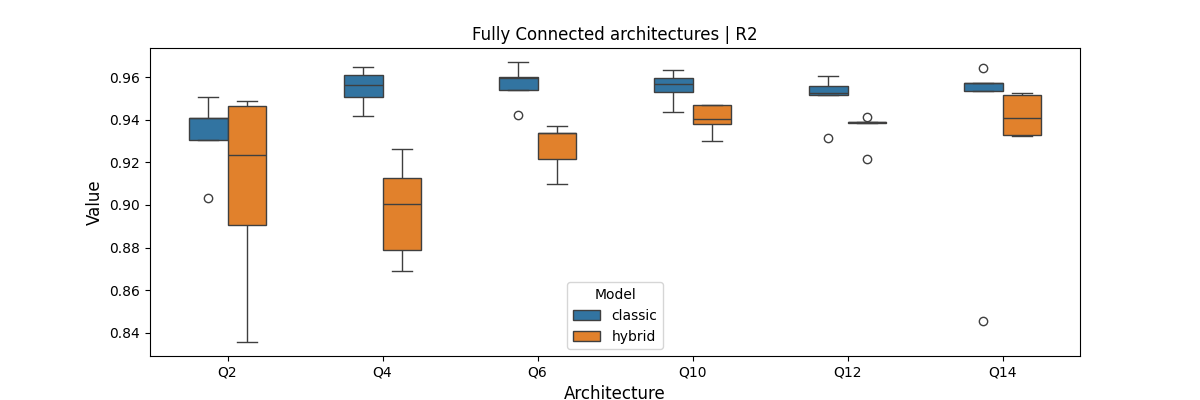}
    \caption{
     Performance scores of all fully connected models explored in this study, as evaluated on the 5 test (hold-out) folds of the CV.
     From \textit{top} to \textit{bottom}: Mean Squared Error (MSE), Mean Absolute Error (MAE), and $R^2$.
     The boxes represent the interquartile range (IQR), which is the range between the 25\textit{th} and 75\textit{th} percentiles of the data.
     The horizontal line inside the box indicates the median value.
     The whiskers extend from the edges of the box to a distance of 1.5 times the IQR, with outliers represented by empty circles.
     Corresponding \textit{classic} (blue)--\textit{hybrid} (orange) models are reported next to each other and indexed by the number of qubits (Q$\langle N_q \rangle$).   
     \label{figure:scores_FC}  
    }
\end{figure*}

\begin{figure*}[htbp]
    \centering
    \includegraphics[width=\textwidth]{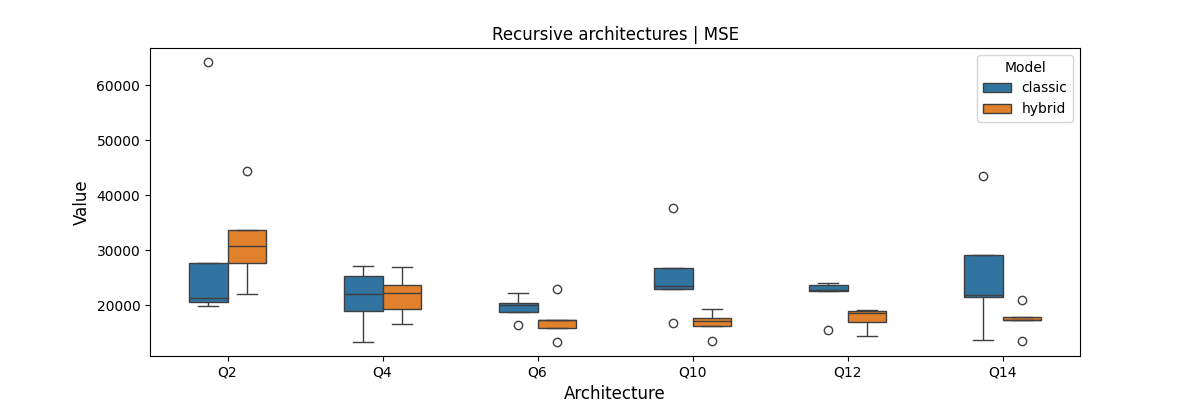}
    \includegraphics[width=\textwidth]{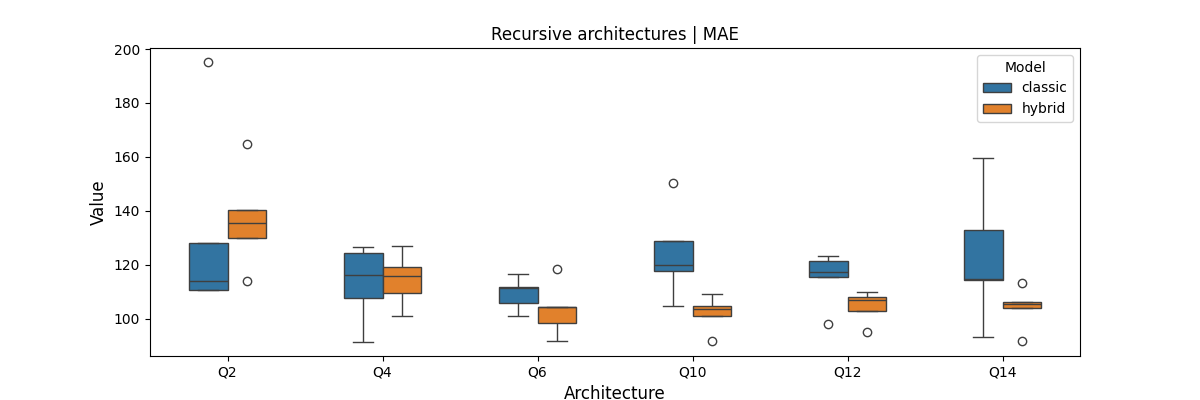}
    \includegraphics[width=\textwidth]{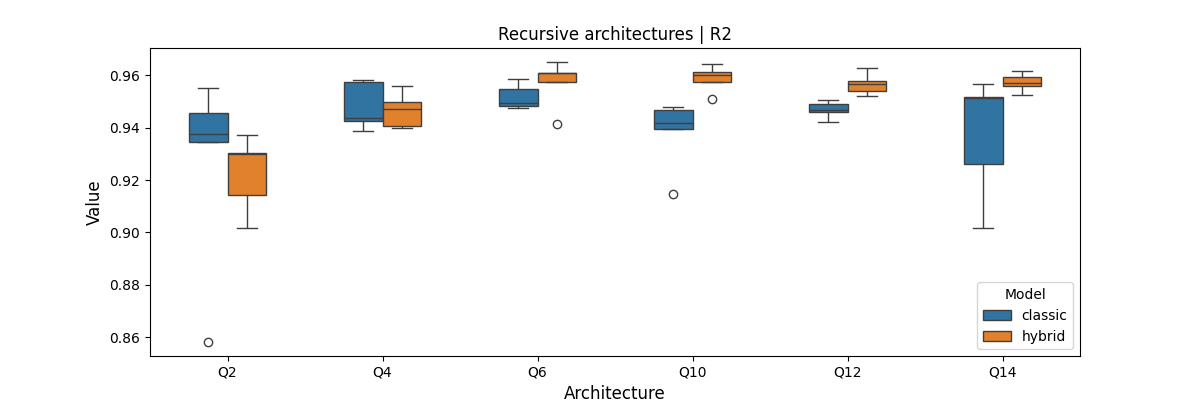}
    \caption{
     Performance scores of all recursive models explored in this study, as evaluated on the 5 test (hold-out) folds of the CV.
     Same as for Figure \ref{figure:scores_FC}.   
     \label{figure:scores_RNN}  
    }
\end{figure*}
As assessment metrics, we considered a collection of commonly adopted regression metrics, namely MSE (also used as a loss function, see Section \ref{Training}), Mean Absolute Error (MAE), and $R^2$ (coefficient of determination).

Figure \ref{figure:scores_FC} shows the measurements relative to the different metrics, estimated over the test folds of the CV loop, for all the fully connected architectures (Scenario A).
From the figure it emerges that, independently of the metric, the discrepancy between classical and hybrid architectures becomes more significant as the number of qubits(/neurons) decreases (except for the case of Q2, which is arguably dominated by outliers, as demonstrated by the sudden drop of performance even for the classic model).
In particular, we observe that the classic/hybrid scores start to be compatible, i.e.\ within their respective uncertainties, from 10 qubits onward.

Figure \ref{figure:scores_RNN} shows the same measurements, this time estimated for the recursive architectures (Scenario B).
In this case, we observe that the hybrid architectures outperform the classical counterparts for 6 qubits or more; moreover, they present lower dispersions, indicating that they are more consistent with respect to the variation of the train/test sets.




\subsection{Consistency Check}
\label{Consistency Check}

\begin{figure}[htbp]
\centering
  \includegraphics[width=0.7\textwidth]{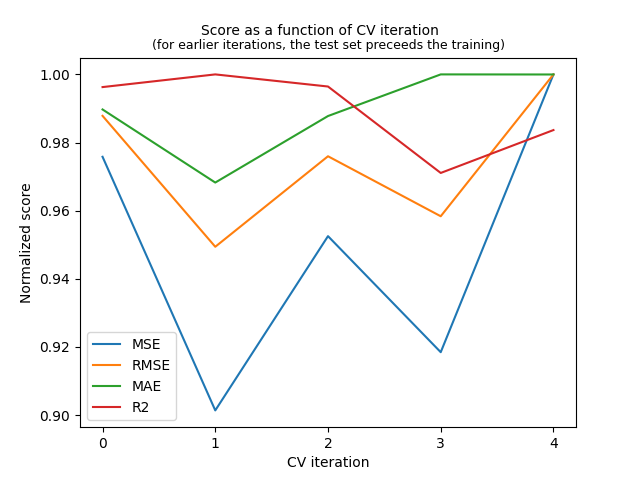}
 \caption{Assessment of a collection of metrics on the test sets belonging to different folding of the cross-validation protocol illustrated in section Section \ref{Performance Evaluation}}
\label{figure:score_vs_cv}
\end{figure}

To rule out any potential bias introduced by our CV assessment protocol, which involved test sets preceding the training sets (as discussed in Section \ref{Performance Evaluation}), we conducted an additional test.
We evaluated the performance scores as a function of the test set's position relative to the training set, as shown in Figure \ref{figure:score_vs_cv}.
Note that, despite this specific plot refers to the hybrid model Q6, it is representative of the results we obtained for all the other models.
The scores have been normalized to 1 for visualization purposes; our interest lies primarily in the trend rather than the absolute values.

Intuitively, Figure \ref{figure:score_vs_cv} explores whether `peeking into the future' -- i.e., placing the test set before the training set -- results in better performances, as would be anticipated if a bias was indeed present.
In such a case we would expect a decline in performance from left to right on the plot, since the GapKFold protocol positions the test set before the training set in earlier iterations (and gradually shifts it to the end in later iterations).
The figure indicates instead that the scores are largely independent of the iteration loop, oscillating around an average value.
This allows us to rule out any significant flaws in our assessment protocol.

\section{Discussion}
\label{Discussion}

The most immediate result of our investigation -- as it readily emerges by the comparison between the left and right panels of Figure \ref{figure:histories}, and the comparison between Figures \ref{figure:scores_FC} and \ref{figure:scores_RNN} -- is that quantum-based NNs do not appear convenient in fully connected architectures, but do offer advantages in recursive architectures (i.e., when data re-upload is involved).
Granted, this conclusion is valid under the assumptions of architecture equivalence defined in Sections \ref{Scenario A} and \ref{Scenario B}.

Surprisingly, despite data re-uploading having operational similarities with RNNs (see Section \ref{Scenario B}) $-$ an industry-standard in time series analysis (see, e.g., \cite{Yu:2023} for quantum-LSTMs) $-$ there is little literature exploring its application in this context.
To the best of our knowledge, \cite{Sagingalieva:2023} is the only work applying data re-upload on a pragmatic time series example.
In this work, the authors study a hybrid encoder-decoder and show that it is superior, in terms of performance, to its classical `equivalent'.
In \cite{Sagingalieva:2023}, the classical equivalent is a Seq2Seq architecture \cite{Sutskever:2014}, while the hybrid model is composed by the same Seq2Seq extended by a quantum circuit including data re-upload.

In our Scenario B approach, we expand on their methodology by \textit{a}) basing the comparison on recursive loops rather than network shape, and \textit{b}) keeping the classical component (in our case, the LSTM autoencoder) as small as possible, using only the encoder and solely for data compression. These improvements allowed us to focus more clearly on the individual contribution of the data re-upload layer.
Additionally, we explored a range of data re-upload iterations (from 2 to 14), emphasizing the novel exploitation of this technique: its usage as a memory cell $-$ rather than merely as a method to enhance a layer's expressivity, as it has so far been employed in regression/classification tasks.

\medskip
Secondarily, we observe that -- regardless of the comparison scenario -- the relative performance of the classic NNs with respect to their hybrid counterparts rapidly declines as network size increases.
Specifically, as the number of qubits increases, the hybrid NNs in fully connected scenarios catch up with the classical ones (Figure \ref{figure:scores_FC}), whereas in recursive architectures, the hybrid NNs progressively diverge towards better performances (Figure \ref{figure:scores_RNN}).
Moreover, we generally observe that this relative performance is non-linear with respect to the number of qubits, evolving more rapidly for lower qubit counts.

\medskip
Adding to these considerations, we draw again attention to how the training curves of Figure \ref{figure:histories} all flatten out before the limiting 20 epochs are reached.
This provides an indication that the aforementioned performance differences are not due to a lack of convergence but are \textit{indeed} related to the models' generalization prowess.
However, focusing on the fully connected architectures (Scenario A; \textit{left} panel) we discover that, while the classical models basically converge within 3-4 epochs, the hybrid models take progressively longer for decreasing qubits counts.
This further suggests that the issue relies on an inferior modeling power of hybrid fully connected architectures.
In contrast, the hybrid architectures exploiting data re-upload (Scenario B; \textit{right} panel) seem to provide the ideal trade, as the models converge within 5 epochs while reaching comparable or even better performances than the fully connected hybrid NNs. 

It is worth noting that these figures are based on simulations of quantum neural networks on conventional hardware, where data re-upload is significantly more computationally expensive.
In our experiments, training a quantum model with $N_q$ qubits took approximately two times longer for each data re-uploading block involved (e.g., \cite{schetakis2024quantum}).
Nevertheless, our results offer valuable insights into the resource demands and optimization strategies necessary when implementing quantum algorithms on actual quantum computers.

\section{Conclusion and Future Directions}\label{conclusion}


In this paper, we present one of the pioneering efforts to explore and quantify the impact of quantum computing and quantum machine learning on traffic forecasting. Our results show that a relatively simple quantum-hybrid model can achieve performance levels comparable to an optimized classical LSTM-based network, though with higher computational complexity.
Most importantly, our findings arguably provide one of the first pieces of evidence suggesting that more sophisticated quantum models, particularly those incorporating data re-uploading, have the potential to surpass classical approaches in terms of performance in the field of time series analysis.
Due to current hardware limitations, we could not fully evaluate these more advanced quantum models for a large number of qubits, but exploring their performance remains a key focus for future research.

\medskip
This work lays the groundwork for the application of Quantum Machine Learning in traffic forecasting and other areas of transportation engineering. The methodologies and insights derived from this study pave the way for further exploration and optimization of quantum-based models for a range of transportation-related challenges.
In these regards, we identify several important avenues to extend our studies in the future. For instance, quantum systems are inherently sensitive to errors and noise, which pose significant challenges to the scalability of quantum computers. This sensitivity impacts both qubits and quantum gates, leading to potential inaccuracies in computations. To study the impact of noise on quantum systems, researchers often use models such as the Lindblad master equation\cite{scully1997quantum} or Kraus operators\cite{nielsen2010quantum}. The Lindblad equation describes the dynamics of open quantum systems under the Born-Markov approximation, while Kraus operators provide a general framework for representing noise channels affecting density matrices.  In simple terms, a noise channel is a linear map that transforms density matrices to other set of density matrices, capturing the effects of decoherence and imperfections in quantum systems.
Since most of our codes have been written using the open-source package PennyLane\cite{bergholm2018pennylane}, we provide general remarks on how to implement noise within this framework. Pennylane offers several methods for simulating noise in quantum circuits. These include classical parametric randomness, the built-in \texttt{default.mixed} device, and plugins for interfacing with other platforms like Cirq and Qiskit. These tools allow researchers to explore how noise affects quantum algorithms and to develop strategies for mitigating its impact.
In summary, gaining a deeper understanding of how noise models affect quantum algorithms can lead to the development of more resilient quantum computing solutions, enhancing their practical applicability across various domains.

An other aspect which offers margin for improvement regards the optimization of learning gradients in non-convex problem structures, an open challenge in quantum machine learning with no established solutions. 
One of the most significant hurdles is the so-called ``Barren plateau''\cite{larocca2024review}, which refers to the issue of vanishing gradients in the optimization landscape of quantum circuits, particularly when scaling to larger system size. Currently, the only types of QNN known to be free from this issue are quantum convolutional neural networks. In light of our current study, we would like to stress that great training results obtained through data re-uploading can be attributed to the fact that data re-uploading variational quantum circuits are, in some instances, barren plateau-free. Effectively, these circuits allow to escape from local minima by repeatedly re-introducing data into the network ~\cite{barthe2024gradients,coelho2024vqc}. For instance, an empirical study conducted in \cite{coelho2024vqc} demonstrated that the magnitude and variance of the gradients remain substantial throughout training, even as the number of qubits increases. This behavior contrasts with the typical degradation seen in barren plateaus.
Alternative approaches to addressing the vanishing gradient issue include characterizing the loss function landscape through Hessian computation, which can provide valuable insights \cite{cerezo2021higher,sen2022variational}. By analyzing the eigenvalues of the Hessian, which quantify local curvature, it is possible to adapt the learning rate to achieve faster convergence during training. These techniques hold promise for mitigating optimization challenges and improving the efficiency of quantum training processes.

\medskip
Looking ahead, our future research will focus on more generalizable comparisons between classical and quantum approaches, particularly in more complex scenarios involving difficult datasets with fewer observations, as well as multi-input and multi-output configurations. Additionally, we plan to assess both approaches in terms of generalizability and transferability, employing a more inclusive and stringent assessment protocol to fully understand the capabilities and limitations of Quantum Machine Learning in real-world applications.


\begin{thebibliography}{10}
\urlstyle{rm}
\expandafter\ifx\csname url\endcsname\relax
  \def\url#1{\texttt{#1}}\fi
\expandafter\ifx\csname urlprefix\endcsname\relax\def\urlprefix{URL }\fi
\expandafter\ifx\csname doiprefix\endcsname\relax\def\doiprefix{DOI: }\fi
\providecommand{\bibinfo}[2]{#2}
\providecommand{\eprint}[2][]{\url{#2}}

\bibitem{Yin:2021}
\bibinfo{author}{Yin, X.} \emph{et~al.}
\newblock \bibinfo{journal}{\bibinfo{title}{Deep learning on traffic prediction: Methods, analysis and future directions}}.
\newblock {\emph{\JournalTitle{IEEE Transactions on Intelligent Transportation Systems}}} \bibinfo{pages}{1--15}, \doiprefix\url{10.1109/TITS.2021.3054840} (\bibinfo{year}{2021}).

\bibitem{Lana:2018}
\bibinfo{author}{Lana, I.}, \bibinfo{author}{Del~Ser, J.}, \bibinfo{author}{Velez, M.} \& \bibinfo{author}{Vlahogianni, E.~I.}
\newblock \bibinfo{journal}{\bibinfo{title}{Road traffic forecasting: Recent advances and new challenges}}.
\newblock {\emph{\JournalTitle{IEEE Intelligent Transportation Systems Magazine}}} \textbf{\bibinfo{volume}{10}}, \bibinfo{pages}{93--109}, \doiprefix\url{10.1109/MITS.2018.2806634} (\bibinfo{year}{2018}).

\bibitem{Cheng:2020}
\bibinfo{author}{Cheng, Z.}, \bibinfo{author}{Pang, M.~S.} \& \bibinfo{author}{Pavlou, P.~A.}
\newblock \bibinfo{journal}{\bibinfo{title}{Mitigating traffic congestion: The role of intelligent transportation systems}}.
\newblock {\emph{\JournalTitle{Information Systems Research}}} \textbf{\bibinfo{volume}{31}}, \bibinfo{pages}{653--674} (\bibinfo{year}{2020}).

\bibitem{Gao:2022}
\bibinfo{author}{Gao, Y.}, \bibinfo{author}{Zhou, C.}, \bibinfo{author}{Rong, J.}, \bibinfo{author}{Wang, Y.} \& \bibinfo{author}{Liu, S.}
\newblock \bibinfo{journal}{\bibinfo{title}{Short-term traffic speed forecasting using a deep learning method based on multitemporal traffic flow volume}}.
\newblock {\emph{\JournalTitle{IEEE Access}}} \textbf{\bibinfo{volume}{10}}, \bibinfo{pages}{82384--82395} (\bibinfo{year}{2022}).

\bibitem{Vlahogianni:2014}
\bibinfo{author}{Vlahogianni, E.~I.}, \bibinfo{author}{Karlaftis, M.~G.} \& \bibinfo{author}{Golias, J.~C.}
\newblock \bibinfo{journal}{\bibinfo{title}{Short-term traffic forecasting: Where we are and where we’re going}}.
\newblock {\emph{\JournalTitle{Transportation Research Part C: Emerging Technologies}}} \textbf{\bibinfo{volume}{43}}, \bibinfo{pages}{3--19}, \doiprefix\url{10.1016/j.trc.2014.01.005} (\bibinfo{year}{2014}).

\bibitem{Yao:2019}
\bibinfo{author}{Yao, H.}, \bibinfo{author}{Tang, X.}, \bibinfo{author}{Wei, H.}, \bibinfo{author}{Zheng, G.} \& \bibinfo{author}{Li, Z.}
\newblock \bibinfo{title}{Revisiting spatial-temporal similarity: A deep learning framework for traffic prediction}.
\newblock In \emph{\bibinfo{booktitle}{Proceedings of the AAAI Conference on Artificial Intelligence}}, vol.~\bibinfo{volume}{33}, \bibinfo{pages}{5668--5675}, \doiprefix\url{10.1609/aaai.v33i01.33015668} (\bibinfo{year}{2019}).

\bibitem{Mantouka:2021}
\bibinfo{author}{Mantouka, E.}, \bibinfo{author}{Barmpounakis, E.}, \bibinfo{author}{Vlahogianni, E.} \& \bibinfo{author}{Golias, J.}
\newblock \bibinfo{journal}{\bibinfo{title}{Smartphone sensing for understanding driving behavior: Current practice and challenges}}.
\newblock {\emph{\JournalTitle{International Journal of Transportation Science and Technology}}} \textbf{\bibinfo{volume}{10}}, \bibinfo{pages}{266--282}, \doiprefix\url{10.1016/j.ijtst.2020.07.001} (\bibinfo{year}{2021}).

\bibitem{Vlahogianni:2017}
\bibinfo{author}{Vlahogianni, E.~I.} \& \bibinfo{author}{Barmpounakis, E.~N.}
\newblock \bibinfo{journal}{\bibinfo{title}{Driving analytics using smartphones: Algorithms, comparisons and challenges}}.
\newblock {\emph{\JournalTitle{Transportation Research Part C: Emerging Technologies}}} \textbf{\bibinfo{volume}{79}}, \bibinfo{pages}{196--206}, \doiprefix\url{10.1016/j.trc.2017.03.014} (\bibinfo{year}{2017}).

\bibitem{Jiang:2021}
\bibinfo{author}{Jiang, W.} \& \bibinfo{author}{Luo, J.}
\newblock \bibinfo{journal}{\bibinfo{title}{Graph neural network for traffic forecasting: A survey}}.
\newblock {\emph{\JournalTitle{IEEE Transactions on Intelligent Transportation Systems}}}  (\bibinfo{year}{2021}).

\bibitem{bharti2022noisy}
\bibinfo{author}{Bharti, K.} \emph{et~al.}
\newblock \bibinfo{journal}{\bibinfo{title}{Noisy intermediate-scale quantum algorithms}}.
\newblock {\emph{\JournalTitle{Reviews of Modern Physics}}} \textbf{\bibinfo{volume}{94}}, \bibinfo{pages}{015004} (\bibinfo{year}{2022}).

\bibitem{preskill2018quantum}
\bibinfo{author}{Preskill, J.}
\newblock \bibinfo{journal}{\bibinfo{title}{Quantum computing in the nisq era and beyond}}.
\newblock {\emph{\JournalTitle{Quantum}}} \textbf{\bibinfo{volume}{2}}, \bibinfo{pages}{79} (\bibinfo{year}{2018}).

\bibitem{arute2019quantum}
\bibinfo{author}{Arute, F.} \emph{et~al.}
\newblock \bibinfo{journal}{\bibinfo{title}{Quantum supremacy using a programmable superconducting processor}}.
\newblock {\emph{\JournalTitle{Nature}}} \textbf{\bibinfo{volume}{574}}, \bibinfo{pages}{505--510} (\bibinfo{year}{2019}).

\bibitem{harrow2017quantum}
\bibinfo{author}{Harrow, A.~W.} \& \bibinfo{author}{Montanaro, A.}
\newblock \bibinfo{journal}{\bibinfo{title}{Quantum computational supremacy}}.
\newblock {\emph{\JournalTitle{Nature}}} \textbf{\bibinfo{volume}{549}}, \bibinfo{pages}{203--209} (\bibinfo{year}{2017}).

\bibitem{zhong2020quantum}
\bibinfo{author}{Zhong, H.-S.} \emph{et~al.}
\newblock \bibinfo{journal}{\bibinfo{title}{Quantum computational advantage using photons}}.
\newblock {\emph{\JournalTitle{Science}}} \textbf{\bibinfo{volume}{370}}, \bibinfo{pages}{1460--1463} (\bibinfo{year}{2020}).

\bibitem{bluvstein2022quantum}
\bibinfo{author}{Bluvstein, D.} \emph{et~al.}
\newblock \bibinfo{journal}{\bibinfo{title}{A quantum processor based on coherent transport of entangled atom arrays}}.
\newblock {\emph{\JournalTitle{Nature}}} \textbf{\bibinfo{volume}{604}}, \bibinfo{pages}{451--456} (\bibinfo{year}{2022}).

\bibitem{bravyi2024high}
\bibinfo{author}{Bravyi, S.} \emph{et~al.}
\newblock \bibinfo{journal}{\bibinfo{title}{High-threshold and low-overhead fault-tolerant quantum memory}}.
\newblock {\emph{\JournalTitle{Nature}}} \textbf{\bibinfo{volume}{627}}, \bibinfo{pages}{778--782} (\bibinfo{year}{2024}).

\bibitem{biamonte2017quantum}
\bibinfo{author}{Biamonte, J.} \emph{et~al.}
\newblock \bibinfo{journal}{\bibinfo{title}{Quantum machine learning}}.
\newblock {\emph{\JournalTitle{Nature}}} \textbf{\bibinfo{volume}{549}}, \bibinfo{pages}{195--202} (\bibinfo{year}{2017}).

\bibitem{cerezo2022challenges}
\bibinfo{author}{Cerezo, M.}, \bibinfo{author}{Verdon, G.}, \bibinfo{author}{Huang, H.-Y.}, \bibinfo{author}{Cincio, L.} \& \bibinfo{author}{Coles, P.~J.}
\newblock \bibinfo{journal}{\bibinfo{title}{Challenges and opportunities in quantum machine learning}}.
\newblock {\emph{\JournalTitle{Nature Computational Science}}} \textbf{\bibinfo{volume}{2}}, \bibinfo{pages}{567--576} (\bibinfo{year}{2022}).

\bibitem{bowles2024better}
\bibinfo{author}{Bowles, J.}, \bibinfo{author}{Ahmed, S.} \& \bibinfo{author}{Schuld, M.}
\newblock \bibinfo{journal}{\bibinfo{title}{Better than classical? the subtle art of benchmarking quantum machine learning models}}.
\newblock {\emph{\JournalTitle{arXiv preprint arXiv:2403.07059}}}  (\bibinfo{year}{2024}).

\bibitem{harrow2009quantum}
\bibinfo{author}{Harrow, A.~W.}, \bibinfo{author}{Hassidim, A.} \& \bibinfo{author}{Lloyd, S.}
\newblock \bibinfo{journal}{\bibinfo{title}{Quantum algorithm for linear systems of equations}}.
\newblock {\emph{\JournalTitle{Physical review letters}}} \textbf{\bibinfo{volume}{103}}, \bibinfo{pages}{150502} (\bibinfo{year}{2009}).

\bibitem{huang2019near}
\bibinfo{author}{Huang, H.-Y.}, \bibinfo{author}{Bharti, K.} \& \bibinfo{author}{Rebentrost, P.}
\newblock \bibinfo{journal}{\bibinfo{title}{Near-term quantum algorithms for linear systems of equations}}.
\newblock {\emph{\JournalTitle{arXiv preprint arXiv:1909.07344}}}  (\bibinfo{year}{2019}).

\bibitem{morales2024quantum}
\bibinfo{author}{Morales, M.~E.} \emph{et~al.}
\newblock \bibinfo{journal}{\bibinfo{title}{Quantum linear system solvers: A survey of algorithms and applications}}.
\newblock {\emph{\JournalTitle{arXiv preprint arXiv:2411.02522}}}  (\bibinfo{year}{2024}).

\bibitem{rebentrost2018quantum}
\bibinfo{author}{Rebentrost, P.}, \bibinfo{author}{Steffens, A.}, \bibinfo{author}{Marvian, I.} \& \bibinfo{author}{Lloyd, S.}
\newblock \bibinfo{journal}{\bibinfo{title}{Quantum singular-value decomposition of nonsparse low-rank matrices}}.
\newblock {\emph{\JournalTitle{Physical review A}}} \textbf{\bibinfo{volume}{97}}, \bibinfo{pages}{012327} (\bibinfo{year}{2018}).

\bibitem{lloyd2014quantum}
\bibinfo{author}{Lloyd, S.}, \bibinfo{author}{Mohseni, M.} \& \bibinfo{author}{Rebentrost, P.}
\newblock \bibinfo{journal}{\bibinfo{title}{Quantum principal component analysis}}.
\newblock {\emph{\JournalTitle{Nature physics}}} \textbf{\bibinfo{volume}{10}}, \bibinfo{pages}{631--633} (\bibinfo{year}{2014}).

\bibitem{Khan:2020}
\bibinfo{author}{Khan, T.~M.} \& \bibinfo{author}{Robles-Kelly, A.}
\newblock \bibinfo{journal}{\bibinfo{title}{Machine learning: Quantum vs classical}}.
\newblock {\emph{\JournalTitle{IEEE Access}}} \textbf{\bibinfo{volume}{8}}, \bibinfo{pages}{219275--219294} (\bibinfo{year}{2020}).

\bibitem{RiveraRuiz:2022}
\bibinfo{author}{Rivera-Ruiz, M.~A.}, \bibinfo{author}{Mendez-Vazquez, A.} \& \bibinfo{author}{López-Romero, J.~M.}
\newblock \bibinfo{title}{Time series forecasting with quantum machine learning architectures}.
\newblock In \emph{\bibinfo{booktitle}{Mexican International Conference on Artificial Intelligence}}, \bibinfo{pages}{66--82} (\bibinfo{publisher}{Springer Nature}, \bibinfo{address}{Cham, Switzerland}, \bibinfo{year}{2022}).

\bibitem{EasomMccaldin:2021}
\bibinfo{author}{Easom-Mccaldin, P.}, \bibinfo{author}{Bouridane, A.}, \bibinfo{author}{Belatreche, A.} \& \bibinfo{author}{Jiang, R.}
\newblock \bibinfo{journal}{\bibinfo{title}{On depth, robustness and performance using the data re-uploading single-qubit classifier}}.
\newblock {\emph{\JournalTitle{IEEE Access}}} \textbf{\bibinfo{volume}{9}}, \bibinfo{pages}{65127--65139} (\bibinfo{year}{2021}).

\bibitem{perez2020data}
\bibinfo{author}{P{\'e}rez-Salinas, A.}, \bibinfo{author}{Cervera-Lierta, A.}, \bibinfo{author}{Gil-Fuster, E.} \& \bibinfo{author}{Latorre, J.~I.}
\newblock \bibinfo{journal}{\bibinfo{title}{Data re-uploading for a universal quantum classifier}}.
\newblock {\emph{\JournalTitle{Quantum}}} \textbf{\bibinfo{volume}{4}}, \bibinfo{pages}{226} (\bibinfo{year}{2020}).

\bibitem{perez2021one}
\bibinfo{author}{P{\'e}rez-Salinas, A.}, \bibinfo{author}{L{\'o}pez-N{\'u}{\~n}ez, D.}, \bibinfo{author}{Garc{\'\i}a-S{\'a}ez, A.}, \bibinfo{author}{Forn-D{\'\i}az, P.} \& \bibinfo{author}{Latorre, J.~I.}
\newblock \bibinfo{journal}{\bibinfo{title}{One qubit as a universal approximant}}.
\newblock {\emph{\JournalTitle{Physical Review A}}} \textbf{\bibinfo{volume}{104}}, \bibinfo{pages}{012405} (\bibinfo{year}{2021}).

\bibitem{Zhao:2017}
\bibinfo{author}{Zhao, Z.}, \bibinfo{author}{Chen, W.}, \bibinfo{author}{Wu, X.}, \bibinfo{author}{Chen, P. C.~Y.} \& \bibinfo{author}{Liu, J.}
\newblock \bibinfo{journal}{\bibinfo{title}{Lstm network: A deep learning approach for short-term traffic forecast}}.
\newblock {\emph{\JournalTitle{IET Intelligent Transportation Systems}}} \textbf{\bibinfo{volume}{11}}, \bibinfo{pages}{68--75}, \doiprefix\url{10.1049/iet-its.2016.0208} (\bibinfo{year}{2017}).

\bibitem{Bapaume:2021}
\bibinfo{author}{Bapaume, T.}, \bibinfo{author}{Côme, E.}, \bibinfo{author}{Roos, J.}, \bibinfo{author}{Ameli, M.} \& \bibinfo{author}{Oukhellou, L.}
\newblock \bibinfo{journal}{\bibinfo{title}{Image inpainting and deep learning to forecast short-term train loads}}.
\newblock {\emph{\JournalTitle{IEEE Access}}} \textbf{\bibinfo{volume}{9}}, \bibinfo{pages}{98506--98522} (\bibinfo{year}{2021}).

\bibitem{Wang:2019}
\bibinfo{author}{Wang, Y.}, \bibinfo{author}{Zhang, D.}, \bibinfo{author}{Liu, Y.}, \bibinfo{author}{Dai, B.} \& \bibinfo{author}{Lee, L.~H.}
\newblock \bibinfo{journal}{\bibinfo{title}{Enhancing transportation systems via deep learning: A survey}}.
\newblock {\emph{\JournalTitle{Transportation Research Part C: Emerging Technologies}}} \textbf{\bibinfo{volume}{99}}, \bibinfo{pages}{144--163}, \doiprefix\url{10.1016/j.trc.2018.12.004} (\bibinfo{year}{2019}).

\bibitem{Fafoutellis:2020}
\bibinfo{author}{Fafoutellis, P.}, \bibinfo{author}{Vlahogianni, E.~I.} \& \bibinfo{author}{Del~Ser, J.}
\newblock \bibinfo{title}{Dilated lstm networks for short-term traffic forecasting using network-wide vehicle trajectory data}.
\newblock In \emph{\bibinfo{booktitle}{Proceedings of the 2020 IEEE 23rd International Conference on Intelligent Transportation Systems (ITSC)}}, \doiprefix\url{10.1109/ITSC45102.2020.9294752} (\bibinfo{year}{2020}).

\bibitem{Ranjan:2020}
\bibinfo{author}{Ranjan, N.}, \bibinfo{author}{Bhandari, S.}, \bibinfo{author}{Zhao, H.~P.}, \bibinfo{author}{Kim, H.} \& \bibinfo{author}{Khan, P.}
\newblock \bibinfo{journal}{\bibinfo{title}{City-wide traffic congestion prediction based on cnn, lstm and transpose cnn}}.
\newblock {\emph{\JournalTitle{IEEE Access}}} \textbf{\bibinfo{volume}{8}}, \bibinfo{pages}{81606--81620} (\bibinfo{year}{2020}).

\bibitem{Boukerche:2020}
\bibinfo{author}{Boukerche, A.} \& \bibinfo{author}{Wang, J.}
\newblock \bibinfo{journal}{\bibinfo{title}{Machine learning-based traffic prediction models for intelligent transportation systems}}.
\newblock {\emph{\JournalTitle{Computer Networks}}} \textbf{\bibinfo{volume}{181}}, \bibinfo{pages}{107530}, \doiprefix\url{10.1016/j.comnet.2020.107530} (\bibinfo{year}{2020}).

\bibitem{Ma:2017}
\bibinfo{author}{Ma, X.} \emph{et~al.}
\newblock \bibinfo{journal}{\bibinfo{title}{Learning traffic as images: A deep convolutional neural network for large-scale transportation network speed prediction}}.
\newblock {\emph{\JournalTitle{Sensors}}} \textbf{\bibinfo{volume}{17}}, \doiprefix\url{10.3390/s17040818} (\bibinfo{year}{2017}).

\bibitem{Dai:2019}
\bibinfo{author}{Dai, X.} \emph{et~al.}
\newblock \bibinfo{journal}{\bibinfo{title}{Deeptrend 2.0: A light-weighted multi-scale traffic prediction model using detrending}}.
\newblock {\emph{\JournalTitle{Transportation Research Part C: Emerging Technologies}}} \textbf{\bibinfo{volume}{103}}, \bibinfo{pages}{142--157}, \doiprefix\url{10.1016/J.TRC.2019.03.022} (\bibinfo{year}{2019}).

\bibitem{Leiser:2021}
\bibinfo{author}{Leiser, N.} \& \bibinfo{author}{Yildirimoglu, M.}
\newblock \bibinfo{journal}{\bibinfo{title}{Incorporating congestion patterns into spatio-temporal deep learning algorithms}}.
\newblock {\emph{\JournalTitle{Transportmetrica B}}} \textbf{\bibinfo{volume}{9}}, \bibinfo{pages}{622--640}, \doiprefix\url{10.1080/21680566.2021.1922320} (\bibinfo{year}{2021}).

\bibitem{Cui:2020}
\bibinfo{author}{Cui, Z.}, \bibinfo{author}{Ke, R.}, \bibinfo{author}{Pu, Z.}, \bibinfo{author}{Ma, X.} \& \bibinfo{author}{Wang, Y.}
\newblock \bibinfo{journal}{\bibinfo{title}{Learning traffic as a graph: A gated graph wavelet recurrent neural network for network-scale traffic prediction}}.
\newblock {\emph{\JournalTitle{Transportation Research Part C: Emerging Technologies}}} \textbf{\bibinfo{volume}{115}}, \bibinfo{pages}{102620}, \doiprefix\url{10.1016/j.trc.2020.102620} (\bibinfo{year}{2020}).

\bibitem{Yu:2021}
\bibinfo{author}{Yu, J. J.~Q.}, \bibinfo{author}{Markos, C.} \& \bibinfo{author}{Zhang, S.}
\newblock \bibinfo{journal}{\bibinfo{title}{Long-term urban traffic speed prediction with deep learning on graphs}}.
\newblock {\emph{\JournalTitle{IEEE Transactions on Intelligent Transportation Systems}}} \doiprefix\url{10.1109/TITS.2021.3069234} (\bibinfo{year}{2021}).

\bibitem{Wang:2021}
\bibinfo{author}{Wang, S.}, \bibinfo{author}{Pei, Z.}, \bibinfo{author}{Wang, C.} \& \bibinfo{author}{Wu, J.}
\newblock \bibinfo{journal}{\bibinfo{title}{Shaping the future of the application of quantum computing in intelligent transportation system}}.
\newblock {\emph{\JournalTitle{Intelligent and Converged Networks}}} \textbf{\bibinfo{volume}{2}}, \bibinfo{pages}{259--276} (\bibinfo{year}{2021}).

\bibitem{Bentley:2022}
\bibinfo{author}{Bentley, C.~D.}, \bibinfo{author}{Marsh, S.}, \bibinfo{author}{Carvalho, A.~R.}, \bibinfo{author}{Kilby, P.} \& \bibinfo{author}{Biercuk, M.~J.}
\newblock \bibinfo{journal}{\bibinfo{title}{Quantum computing for transport optimization}}.
\newblock {\emph{\JournalTitle{arXiv preprint arXiv:2206.07313}}}  (\bibinfo{year}{2022}).

\bibitem{Dixit:2023}
\bibinfo{author}{Dixit, V.~V.} \& \bibinfo{author}{Niu, C.}
\newblock \bibinfo{journal}{\bibinfo{title}{Quantum computing for transport network design problems}}.
\newblock {\emph{\JournalTitle{Scientific Reports}}} \textbf{\bibinfo{volume}{13}}, \bibinfo{pages}{12267} (\bibinfo{year}{2023}).

\bibitem{DixitRouting:2023}
\bibinfo{author}{Dixit, V.~V.}, \bibinfo{author}{Niu, C.}, \bibinfo{author}{Rey, D.}, \bibinfo{author}{Waller, S.~T.} \& \bibinfo{author}{Levin, M.~W.}
\newblock \bibinfo{journal}{\bibinfo{title}{Quantum computing to solve scenario-based stochastic time-dependent shortest path routing}}.
\newblock {\emph{\JournalTitle{Transportation Letters}}} \bibinfo{pages}{1--11} (\bibinfo{year}{2023}).

\bibitem{Yarkoni:2020}
\bibinfo{author}{Yarkoni, S.} \emph{et~al.}
\newblock \bibinfo{title}{Quantum shuttle: traffic navigation with quantum computing}.
\newblock In \emph{\bibinfo{booktitle}{Proceedings of the 1st ACM SIGSOFT International Workshop on Architectures and Paradigms for Engineering Quantum Software}}, \bibinfo{pages}{22--30} (\bibinfo{year}{2020}).

\bibitem{Brooks:2019}
\bibinfo{author}{Brooks, M.}
\newblock \bibinfo{journal}{\bibinfo{title}{Beyond quantum supremacy: The hunt for useful quantum computers}}.
\newblock {\emph{\JournalTitle{Nature}}} \textbf{\bibinfo{volume}{574}}, \bibinfo{pages}{19--22} (\bibinfo{year}{2019}).

\bibitem{Cerezo:2022}
\bibinfo{author}{Cerezo, M.}, \bibinfo{author}{Verdon, G.}, \bibinfo{author}{Huang, H.-Y.}, \bibinfo{author}{Cincio, L.} \& \bibinfo{author}{Coles, P.~J.}
\newblock \bibinfo{journal}{\bibinfo{title}{Challenges and opportunities in quantum machine learning}}.
\newblock {\emph{\JournalTitle{Nature Computational Science}}} \textbf{\bibinfo{volume}{2}}, \bibinfo{pages}{567--576}, \doiprefix\url{10.1038/s43588-022-00311-3} (\bibinfo{year}{2022}).

\bibitem{Biamonte:2017}
\bibinfo{author}{Biamonte, J.} \emph{et~al.}
\newblock \bibinfo{journal}{\bibinfo{title}{Quantum machine learning}}.
\newblock {\emph{\JournalTitle{Nature}}} \textbf{\bibinfo{volume}{549}}, \bibinfo{pages}{195--202}, \doiprefix\url{10.1038/nature23474} (\bibinfo{year}{2017}).

\bibitem{Cai:2022}
\bibinfo{author}{Cai, H.}, \bibinfo{author}{Ye, Q.} \& \bibinfo{author}{Deng, D.-L.}
\newblock \bibinfo{journal}{\bibinfo{title}{Sample complexity of learning parametric quantum circuits}}.
\newblock {\emph{\JournalTitle{Quantum Science and Technology}}} \textbf{\bibinfo{volume}{7}}, \bibinfo{pages}{025014}, \doiprefix\url{10.1088/2058-9565/ac4f30} (\bibinfo{year}{2022}).

\bibitem{Haug:2021}
\bibinfo{author}{Haug, T.}, \bibinfo{author}{Bharti, K.} \& \bibinfo{author}{Kim, M.~S.}
\newblock \bibinfo{journal}{\bibinfo{title}{Capacity and quantum geometry of parametrized quantum circuits}}.
\newblock {\emph{\JournalTitle{PRX Quantum}}} \textbf{\bibinfo{volume}{2}}, \bibinfo{pages}{040309}, \doiprefix\url{10.1103/PRXQuantum.2.040309} (\bibinfo{year}{2021}).

\bibitem{Mari:2020}
\bibinfo{author}{Mari, A.}, \bibinfo{author}{Bromley, T.~R.}, \bibinfo{author}{Izaac, J.}, \bibinfo{author}{Schuld, M.} \& \bibinfo{author}{Killoran, N.}
\newblock \bibinfo{journal}{\bibinfo{title}{Transfer learning in hybrid classical-quantum neural networks}}.
\newblock {\emph{\JournalTitle{Quantum}}} \textbf{\bibinfo{volume}{4}}, \bibinfo{pages}{340}, \doiprefix\url{10.22331/q-2020-10-09-340} (\bibinfo{year}{2020}).

\bibitem{Schetakis:2022}
\bibinfo{author}{Schetakis, N.}, \bibinfo{author}{Aghamalyan, D.}, \bibinfo{author}{Griffin, P.} \& \bibinfo{author}{Boguslavsky, M.}
\newblock \bibinfo{journal}{\bibinfo{title}{Review of some existing qml frameworks and novel hybrid classical–quantum neural networks realizing binary classification for noisy datasets}}.
\newblock {\emph{\JournalTitle{Scientific Reports}}} \textbf{\bibinfo{volume}{12}}, \bibinfo{pages}{11927}, \doiprefix\url{10.1038/s41598-022-14876-6} (\bibinfo{year}{2022}).

\bibitem{SchetakisCredit:2022}
\bibinfo{author}{Schetakis, N.} \emph{et~al.}
\newblock \bibinfo{journal}{\bibinfo{title}{Quantum machine learning for credit scoring}}.
\newblock {\emph{\JournalTitle{Research Collection School of Computing and Information Systems}}} \bibinfo{pages}{1--13} (\bibinfo{year}{2022}).

\bibitem{Bharti:2022}
\bibinfo{author}{Bharti, K.} \emph{et~al.}
\newblock \bibinfo{journal}{\bibinfo{title}{Noisy intermediate-scale quantum algorithms}}.
\newblock {\emph{\JournalTitle{Reviews of Modern Physics}}} \textbf{\bibinfo{volume}{94}}, \bibinfo{pages}{015004}, \doiprefix\url{10.1103/RevModPhys.94.015004} (\bibinfo{year}{2022}).

\bibitem{CerezoVQA:2021}
\bibinfo{author}{Cerezo, M.} \emph{et~al.}
\newblock \bibinfo{journal}{\bibinfo{title}{Variational quantum algorithms}}.
\newblock {\emph{\JournalTitle{Nature Reviews Physics}}} \textbf{\bibinfo{volume}{3}}, \bibinfo{pages}{625--644}, \doiprefix\url{10.1038/s42254-021-00348-9} (\bibinfo{year}{2021}).

\bibitem{Schuld:2021}
\bibinfo{author}{Schuld, M.}, \bibinfo{author}{Sweke, R.} \& \bibinfo{author}{Meyer, J.~J.}
\newblock \bibinfo{journal}{\bibinfo{title}{Effect of data encoding on the expressive power of variational quantum-machine-learning models}}.
\newblock {\emph{\JournalTitle{Physical Review A}}} \textbf{\bibinfo{volume}{103}}, \bibinfo{pages}{032430} (\bibinfo{year}{2021}).

\bibitem{Bergholm:2018}
\bibinfo{author}{Bergholm, V.} \emph{et~al.}
\newblock \bibinfo{journal}{\bibinfo{title}{Pennylane: Automatic differentiation of hybrid quantum-classical computations}}.
\newblock {\emph{\JournalTitle{arXiv preprint arXiv:1811.04968}}}  (\bibinfo{year}{2018}).
\newblock \bibinfo{note}{[Online]. Available: http://arxiv.org/abs/1811.04968. Accessed: Jul. 31, 2023.}

\bibitem{Bergmeir:2018}
\bibinfo{author}{Bergmeir, C.}, \bibinfo{author}{Hyndman, R.~J.} \& \bibinfo{author}{Koo, B.}
\newblock \bibinfo{journal}{\bibinfo{title}{A note on the validity of cross-validation for evaluating autoregressive time series prediction}}.
\newblock {\emph{\JournalTitle{Computational Statistics \& Data Analysis}}} \textbf{\bibinfo{volume}{120}}, \bibinfo{pages}{70--83}, \doiprefix\url{10.1016/j.csda.2017.11.003} (\bibinfo{year}{2018}).

\bibitem{Racine:2000}
\bibinfo{author}{Racine, J.}
\newblock \bibinfo{journal}{\bibinfo{title}{Consistent cross-validatory model-selection for dependent data: Hv-block cross-validation}}.
\newblock {\emph{\JournalTitle{Journal of Econometrics}}} \textbf{\bibinfo{volume}{99}}, \bibinfo{pages}{39--61}, \doiprefix\url{10.1016/S0304-4076(00)00030-0} (\bibinfo{year}{2000}).

\bibitem{Zheng:2023}
\bibinfo{author}{Zheng, W.}
\newblock \bibinfo{journal}{\bibinfo{title}{Tscv: Time series cross-validation}}.
\newblock {\emph{\JournalTitle{[Online]. Available: https://arxiv.org/abs/2307.02201}}}  (\bibinfo{year}{2023}).

\bibitem{Yu:2023}
\bibinfo{author}{Yu, Y.}, \bibinfo{author}{Hu, G.}, \bibinfo{author}{Liu, C.}, \bibinfo{author}{Xiong, J.} \& \bibinfo{author}{Wu, Z.}
\newblock \bibinfo{journal}{\bibinfo{title}{Prediction of solar irradiance one hour ahead based on quantum long short-term memory network}}.
\newblock {\emph{\JournalTitle{IEEE Transactions on Quantum Engineering}}} \textbf{\bibinfo{volume}{4}}, \bibinfo{pages}{1--15} (\bibinfo{year}{2023}).

\bibitem{Sagingalieva:2023}
\bibinfo{author}{Sagingalieva, A.} \emph{et~al.}
\newblock \bibinfo{journal}{\bibinfo{title}{Photovoltaic power forecasting using quantum machine learning}}.
\newblock {\emph{\JournalTitle{arXiv preprint arXiv:2312.16379}}}  (\bibinfo{year}{2023}).
\newblock \bibinfo{note}{[Online]. Available: http://arxiv.org/abs/2312.16379}.

\bibitem{Sutskever:2014}
\bibinfo{author}{Sutskever, I.}
\newblock \bibinfo{title}{Sequence to sequence learning with neural networks}.
\newblock In \emph{\bibinfo{booktitle}{Advances in Neural Information Processing Systems 27}} (\bibinfo{year}{2014}).

\bibitem{schetakis2024quantum}
\bibinfo{author}{Schetakis, N.} \emph{et~al.}
\newblock \bibinfo{journal}{\bibinfo{title}{Quantum machine learning for credit scoring}}.
\newblock {\emph{\JournalTitle{Mathematics}}} \textbf{\bibinfo{volume}{12}}, \bibinfo{pages}{1391} (\bibinfo{year}{2024}).

\bibitem{scully1997quantum}
\bibinfo{author}{Scully, M.~O.} \& \bibinfo{author}{Zubairy, M.~S.}
\newblock \emph{\bibinfo{title}{Quantum optics}} (\bibinfo{publisher}{Cambridge university press}, \bibinfo{year}{1997}).

\bibitem{nielsen2010quantum}
\bibinfo{author}{Nielsen, M.~A.} \& \bibinfo{author}{Chuang, I.~L.}
\newblock \emph{\bibinfo{title}{Quantum computation and quantum information}} (\bibinfo{publisher}{Cambridge university press}, \bibinfo{year}{2010}).

\bibitem{bergholm2018pennylane}
\bibinfo{author}{Bergholm, V.} \emph{et~al.}
\newblock \bibinfo{journal}{\bibinfo{title}{Pennylane: Automatic differentiation of hybrid quantum-classical computations}}.
\newblock {\emph{\JournalTitle{arXiv preprint arXiv:1811.04968}}}  (\bibinfo{year}{2018}).

\bibitem{larocca2024review}
\bibinfo{author}{Larocca, M.} \emph{et~al.}
\newblock \bibinfo{journal}{\bibinfo{title}{A review of barren plateaus in variational quantum computing}}.
\newblock {\emph{\JournalTitle{arXiv preprint arXiv:2405.00781}}}  (\bibinfo{year}{2024}).

\bibitem{barthe2024gradients}
\bibinfo{author}{Barthe, A.} \& \bibinfo{author}{P{\'e}rez-Salinas, A.}
\newblock \bibinfo{journal}{\bibinfo{title}{Gradients and frequency profiles of quantum re-uploading models}}.
\newblock {\emph{\JournalTitle{Quantum}}} \textbf{\bibinfo{volume}{8}}, \bibinfo{pages}{1523} (\bibinfo{year}{2024}).

\bibitem{coelho2024vqc}
\bibinfo{author}{Coelho, R.}, \bibinfo{author}{Sequeira, A.} \& \bibinfo{author}{Paulo~Santos, L.}
\newblock \bibinfo{journal}{\bibinfo{title}{Vqc-based reinforcement learning with data re-uploading: performance and trainability}}.
\newblock {\emph{\JournalTitle{Quantum Machine Intelligence}}} \textbf{\bibinfo{volume}{6}}, \bibinfo{pages}{53} (\bibinfo{year}{2024}).

\bibitem{cerezo2021higher}
\bibinfo{author}{Cerezo, M.} \& \bibinfo{author}{Coles, P.~J.}
\newblock \bibinfo{journal}{\bibinfo{title}{Higher order derivatives of quantum neural networks with barren plateaus}}.
\newblock {\emph{\JournalTitle{Quantum Science and Technology}}} \textbf{\bibinfo{volume}{6}}, \bibinfo{pages}{035006} (\bibinfo{year}{2021}).

\bibitem{sen2022variational}
\bibinfo{author}{Sen, P.}, \bibinfo{author}{Bhatia, A.~S.}, \bibinfo{author}{Bhangu, K.~S.} \& \bibinfo{author}{Elbeltagi, A.}
\newblock \bibinfo{journal}{\bibinfo{title}{Variational quantum classifiers through the lens of the hessian}}.
\newblock {\emph{\JournalTitle{Plos one}}} \textbf{\bibinfo{volume}{17}}, \bibinfo{pages}{e0262346} (\bibinfo{year}{2022}).

\end{thebibliography}



\section*{Acknowledgements}

The authors would like to thank the European Union for funding this research through two projects: ERA4CH (Earthquake Risk Platform For European Cities Cultural Heritage Protection—grant agreement No. 101086280) and EYE (Economy bY spacE—grand agreement No. 10100763), both part of the Horizon 2020 research and innovation program. The research was also supported by the Cyprus National Project CODEVELOP-ICT-HEALTH/0322/0047. 

\section*{Author contributions statement}

The authors confirm contribution to the paper as follows: study conception and design: N.S., P.B., N.A., D.A., K.B., P.F., E.I.V.; data collection: P.F., E.I.V.; analysis and interpretation of results: P.B., N.S., K.B., N.A., D.A., S.I.T., A. A., E.I.V.; draft manuscript preparation: N.S., P.B., N.A., P.F.; All authors reviewed the manuscript.




\end{document}